\documentclass[
reprint, twocolumn, secnumarabic%
,amssymb,
nobibnotes, nofootinbib, aps, prl, showpacs,showkeys]{revtex4}
\usepackage{docs}%
\usepackage{bm}%
\usepackage{graphicx}
\usepackage{rotating}
\usepackage{hyperref}
\usepackage{url}
\usepackage{srcltx}
\expandafter \ifx \csname package@font\endcsname\relax\else
\expandafter \expandafter \expandafter
\usepackage
\expandafter \expandafter
\expandafter{\csname package@font\endcsname}%
\fi



\usepackage{enumerate}

\begin{document}

\title{Comparison of Neural Network and Hadronic Model  Predictions of Two-Photon Exchange Effect}
\author{Krzysztof M. Graczyk}

\email{kgraczyk@ift.uni.wroc.pl}
\affiliation{Institute of
Theoretical Physics, University of Wroc\l aw, pl. M. Borna 9,
50-204, Wroc\l aw, Poland}


\begin{abstract}

Predictions for the two-photon exchange (TPE) correction to unpolarized $ep$ elastic cross section, obtained within two different approaches, are confronted and discussed in detail. In the  first one the TPE correction is extracted from  experimental data by applying the Bayesian neural network (BNN) statistical framework. In the other  the TPE is given by  box diagrams, with the nucleon and the $P_{33}$ resonance as the hadronic intermediate states. Two different form factor parametrizations for both the  proton and the $P_{33}$ resonance are taken into consideration.  Proton form factors are obtained from the global  fit of the full model (with the TPE correction) to the unpolarized cross section data. Predictions of both methods  agree well in the intermediate $Q^2$ range,  $(1,3)$ GeV$^2$. Above $Q^2=3$ GeV$^2$ the agreement is on $2\sigma$  level.  Below $Q^2=1$ GeV$^2$ the consistency between both approaches is broken. The values of the proton radius  extracted within both models are given. In both cases  predictions for VEPP-3 experiment  have been obtained and confronted with the preliminary experimental results.

\end{abstract}

\maketitle

\section{Introduction}
\label{Section_Introduction}

\begin{figure}
\centering{
\includegraphics[width=0.4\textwidth]{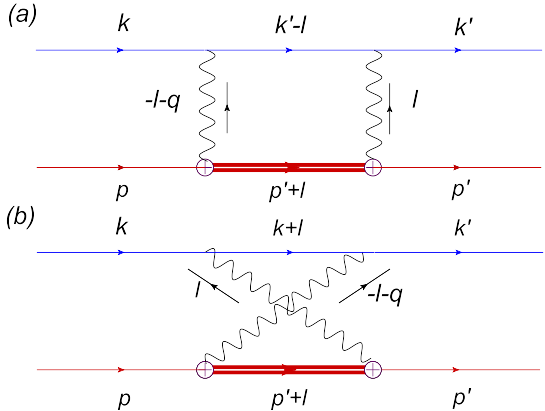}
\caption{(Color online) Direct (a) and exchange (b) TPE box diagrams for elastic $ep$ scattering (Eq. \ref{Fig_boxdiagrms}). The
intermediate states (thick double lines) are given by either the nucleon or the $P_{33}(1232)$
resonance. \label{Fig_boxdiagrms}} }
\end{figure}

Two photon exchange  (TPE) effect in the elastic electron scattering off the proton has drawn back the attention of physicists about ten years ago, when the new experimental technique for measurement of the electromagnetic nucleon form-factors (FFs) had become available.
In this method, called later polarization transfer (PT) technique, various polarization observables are measured and the form factor ratio,
\begin{equation}
\label{FF_ratio}
\mathcal{R}_{1\gamma}(Q^2) = \mu_p \frac{G_E(Q^2)}{ G_M(Q^2)},
\end{equation}
is estimated \cite{Ron:2011rdand_Puckett:2011xg}. $G_E$ and $G_M$ are the electric and the magnetic proton form factors  respectively, $\mu_p=2.793$ is the proton magnetic moment in the units of the nuclear magneton.

The proton electromagnetic FFs are also extracted from  unpolarized cross section data (CS) by applying Rosenbluth (longitudinal-transverse (LT)) separation. As the result the electric and the magnetic FFs are obtained simultaneously.  It turns out that the ratio (\ref{FF_ratio}) estimated basing on the Rosenbluth FFs data  is inconsistent with the PT measurements at larger $Q^2$ values.

It is generally widely accepted that an insufficient estimate of the radiative corrections (RCs) applied in the Rosenbluth data analysis is a main source of inconsistency. In particular, it is argued that a lack of the so-called hard-photon TPE contribution coming from  box diagrams drawn in Fig. \ref{Fig_boxdiagrms} is responsible for the  disagreement\footnote{We notice an  explanation proposed by Bystritskiy et al. \cite{Bystritskiy:2006ju}. } \cite{Blunden:2003sp,Guichon:2003qm,Chen:2004tw}.
\begin{figure}
\centering{
\includegraphics[width=0.52\textwidth]{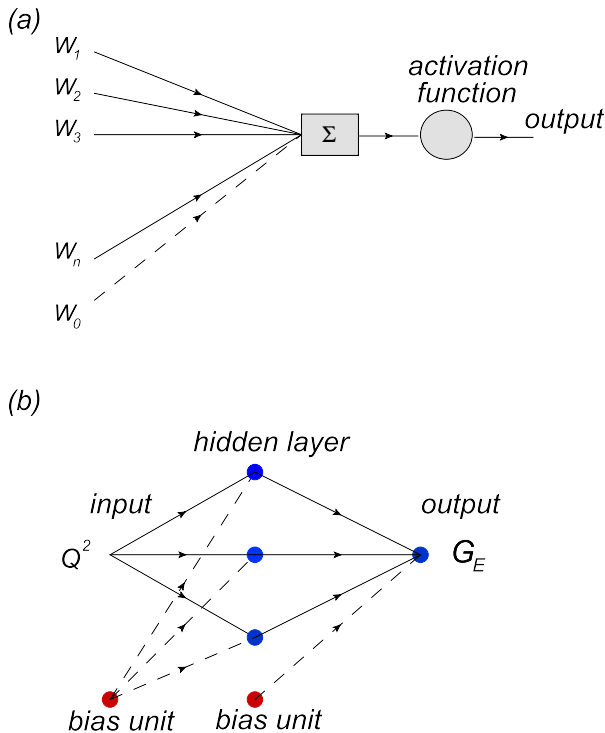}
\caption{(Color online) In the top: single unit connected with $n$ units from the previous layer  and with one bias unit ($\omega_0$ weight).
In the bottom: simple MLP with one hidden layer, used to fit the electric proton form factor data \label{simplenetwork}
}
}
\end{figure}

In the old Rosenbluth data analyses  the cross section measurements were corrected  by the RCs obtained by Mo and Tsai (MT) \cite{MT_radiative}.  In this approach the TPE corrections were calculated within a soft photon approximation.  Supplementing  this contribution by  the hard photon correction changes  the results of the Rosenbluth separation and  makes them nearly consistent with the PT measurements \cite{Blunden:2003sp,Blunden:2005ew}.

Recently several theoretical calculations of the TPE correction have been performed \cite{Blunden:2003sp,Blunden:2005ew,Zhou:2009nf,Kondratyuk:2005kk,Kondratyuk:2006ig,Borisyuk:2006fh,Borisyuk:2008es,Borisyuk:2012he,Chen:2004tw,Afanasev:2005mp,Borisyuk:2008db,Tjon:2009hf,Kivel:2012vs,Borisyuk:2013hja}. They have been done within various approaches (for a review see Refs. \cite{Arrington:2011dn,Carlson:2007sp,Vanderhaeghen:2007zz}). It happens that  the predictions of the TPE effect are mostly model-dependent at larger values of $Q^2$.

Simultaneously to the theoretical activity phenomenological investigations have been carried out as well. An effort has been made to extract the proton FFs and the TPE term directly from the experimental data \cite{Arrington:2003qk,TomasiGustafsson:2004ms,Tvaskis:2005ex,Belushkin:2007zv,Borisyuk:2007re,Arrington:2007ux,Chen:2007ac,Alberico:2008sz,TomasiGustafsson:2009pw,Guttmann:2010au,Venkat:2010by,Graczyk:2011kh,Qattan:2011ke,Qattan:2011zz}.

The TPE effect can be studied experimentally. The TPE correction for the elastic positron-proton scattering has an opposite sign but the same absolute value as the corresponding term in the electron-proton scattering. Hence  the measurement of cross section ratio,
\begin{equation}
\label{R_+_-}
R_{+/-} = \frac{\displaystyle \frac{d\sigma}{d\Omega}(e^+ p \to e^+ p)}{\displaystyle \frac{d\sigma}{d\Omega}(e^- p \to e^- p)} \approx 1 - 2 (\mathrm{TPE}),
\end{equation}
gives a direct possibility to estimate the TPE correction. At present two  dedicated  $R_{+/-} $ measurement experiments are operating \cite{Bennett:2012zza,Gramolin:2011tr}. There is also a proposal of a new project, called OLYMPUS, in DESY \cite{Ice:2012zz}.

In this report we would like to confront the phenomenological estimation  of the TPE effect, obtained in our previous paper \cite{Graczyk:2011kh}, with the theoretical predictions. In Ref. \cite{Graczyk:2011kh} the  global Bayesian analysis of the world elastic $ep$ data was performed.  The major idea was to built a statistical model  based on the experimental measurements with ability to make predictions of the electromagnetic proton FFs and the TPE term. It was achieved by adapting the  Bayesian framework for the feed-forward neural networks (BNN) \cite{Graczyk:2010gw}. This formalism allows one to perform  the analysis as model-independent as possible. However, because the incompleteness  of the data some additional assumptions had to be made.  We applied  constraints coming from   the  $C$-parity and the crossing symmetry invariance of the $ep$ scattering amplitude
\cite{Rekalo:2003km,Rekalo:2003xa,Rekalo:2004wa}. But the most important  was to assume that for the PT data the TPE effect can be neglected and it is only relevant for the cross section data. This statement  is supported both by general  arguments \cite{Guichon:2003qm} and calculations \cite{Blunden:2005ew,Afanasev:2005mp}. Indeed, the TPE corrections to the unpolarized cross section  and the PT ratio $R_{1\gamma}$  data are comparable. But their inclusion in the Rosenbluth analysis  affects significantly the results of the FFs extraction, while in the case of  PT measurements TPE correction is of the  order of  statistical errors.

The comparison of BNN and the theoretical predictions allows one to verify validity of the model assumptions and to confront, in the non-direct way,  theoretical model predictions with the data represented by the BNN.

In this paper the TPE corrections are computed in the similar way as in Refs. \cite{Blunden:2003sp,Zhou:2009nf,Kondratyuk:2005kk,Kondratyuk:2006ig,Blunden:2005ew}, in a quantum-field theory approach, called later as hadronic model (HM).  Hadronic intermediate states in the box diagrams (Fig. \ref{Fig_boxdiagrms}) are given by nucleon  and $P_{33}(1232)$ resonance.   Heavier resonances are not included because it was shown that their total contribution is negligible for these kinematics   \cite{Kondratyuk:2006ig} and their inclusion introduces the additional model-dependence to the discussion.

Our approach should work well at low and intermediate $Q^2$ range.  Its input includes proton and $P_{33}(1232)$ electromagnetic FFs. For the proton we consider two different  types of parametrizations. In contrast to the BNN analysis the FFs parameters are established from the global fit of  HM to the unpolarized cross section data only.  For the $ N \to P_{33}$ electromagnetic transition we consider   different vertex and FFs parametrizations than in Ref. \cite{Kondratyuk:2005kk}.

In a wide $Q^2$ domain the BNN and the hadronic model predictions agree well. The discrepancy appears at low $Q^2$. However, the value of the proton radius extracted  from the BNN fit  is consistent with the recent atomic measurement by Pohl et al. \cite{Pohl:2010zza}. It is shown that the low-$Q^2$ inconsistency between BNN and HM is induced by one of the model  assumptions mentioned  above (neglecting TPE correction to the PT data).

Eventually, we compare the  predictions of the $R_{+/-}$ ratio obtained within the BNN and the HM approaches with the preliminary  VEPP-3 measurements \cite{Gramolin:2011tr}. The theoretical and the phenomenological  predictions are in  agreement with the new available data.

The paper is organized as follows. It contains six sections and two appendixes. In Sec. \ref{Section_Basic_Features}  the basic  formalism is introduced. In Sec. \ref{Section_NN_attempt} Bayesian neural network approach is shortly reviewed. The hadronic model  is described in Sec. \ref{Section_Theoretical_Attempt}.  Detailed comparison of the BNN and the theoretical predictions  are presented in Sec. \ref{Section_Results}. We summarize our results in Sec. \ref{Sec_Summary}.  Some  technical details of the theoretical calculations are enclosed in Appendix \ref{Appendix_A}, while Appendix \ref{Appendix_B} contains the definition of $\chi^2$ function used in the data analysis.
\begin{figure}
\includegraphics[width=0.5\textwidth]{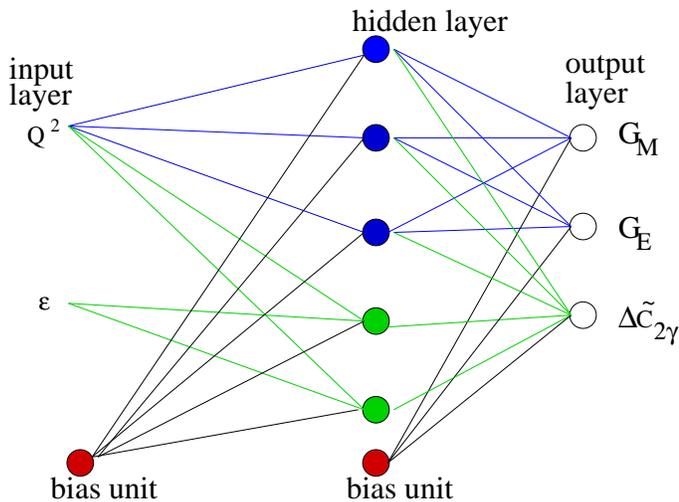}
\caption{(Color online) Network $\mathcal{N}_{3,2}$: two input units, one layer of hidden units, three output units. The FF sector: blue filled units and connections; TPE sector: green units and connections.  Each line corresponds to one weight parameter.  \label{Fig_network}}
\end{figure}

\section{Basic Formalism}
\label{Section_Basic_Features}

We consider  the elastic electron-proton scattering,
\begin{equation}
\label{reaction_ep_to_ep}
e(k) + p(p) \to p(p') + e (k').
\end{equation}
By $k$, $k'$ and $p$, $p'$ the  initial and the final electron's and proton's four-momenta are denoted respectively. A four-momentum transfer  is defined as
$q=k-k'$ and $q^2  = (k-k')^2 =   -Q^2$.

To compute the TPE correction  we apply a  typical approach to
account for the radiative corrections in the $ep$ scattering \cite{Maximon:2000hm,Blunden:2003sp}. It is  the quantum electrodynamics (QED) extended to include the hadronic degrees of freedom, like the proton and the $P_{33}(1232)$ resonance. The proton and nucleon electromagnetic vertices are expressed in terms of transition  FFs.

The matrix element for the $ep$ scattering can be written as  a perturbative series in  $\alpha = e^2/4\pi \approx 1/137$. The first element of the series, $\mathcal{M}_{1\gamma} $, describes an exchange of one photon between the electron and the proton target, and it gives the lowest order contribution of the differential cross section,
$
d\sigma_{1\gamma}   \sim \left|\mathcal{M}_{1\gamma}\right|^2.
$
The $\mathcal{M}_{1\gamma}$ matrix element is a contraction of the one-body
leptonic with hadronic currents,
 \begin{equation}
 i\mathcal{M}_{1\gamma}
 =
 i \frac{e^2}{Q^2} j^\mu h_\mu.
 \end{equation}
The leptonic and hadronic  currents read,
\begin{eqnarray}
\label{current_leptonic_OPE}
j_\mu(q) &=& \overline{u}(k')  \gamma^\mu u(k),\\
\label{current_hadronic_OPE}
h^\mu(q) &=& \overline{u}(p') \Gamma^\mu(q) u(p).
\end{eqnarray}
$\Gamma^\mu$ is the on-shell proton electromagnetic vertex,
\begin{eqnarray}
\label{1gamma_Nphoton_vertex}
\Gamma^\mu(q)  &=& \gamma^\mu F_1(Q^2) + \frac{i\sigma^{\mu\nu} q_\nu}{2M_p} F_2(Q^2).
\end{eqnarray}
$F_1$ and $F_2$ are the Dirac and the spin flip  proton form factors respectively, while $M_p = 938$ MeV$/c^2$ is the proton mass. It is useful to express  the above FFs  in terms of  the electric  and the magnetic  proton FFs.
\begin{eqnarray}
F_1 (Q^2)
&=&
\frac{1}{1+\tau} \left\{ G_E(Q^2) + \tau G_M(Q^2)\right\},
\\
F_2 (Q^2)
&=&
 \frac{1}{1 +\tau} \left\{ G_M(Q^2) -  G_E(Q^2)\right\},
\end{eqnarray}
where
$\tau={Q^2}/{4M^2_{p}}$.

We keep the normalization, $G_E(0)=1$, $G_M(0)=\mu_p$, hence $F_1(0)=1$, $F_2(0)= \kappa_p \equiv \mu_p - 1$.

In the  $ep$ scattering data analysis it is convenient
 to consider the reduced cross section, $\sigma_R$ ($d\sigma/d\Omega \sim \sigma_R$), which in Born approximation is given by the formula,
\begin{equation}
\label{sigma_reduced}
\sigma_{R,1\gamma}(Q^2, \varepsilon) = \tau G_M^2(Q^2) + \varepsilon G_E^2(Q^2),
\end{equation}
where $\varepsilon$ is the photon polarization
\begin{equation}
\varepsilon = \left[1+2\left(1+\frac{Q^2}{4M_p^2}\right)\tan^2\!\left(\frac{\theta}{2}\right)\right]^{-1},
\label{epsilon}
\end{equation}
$\theta$ is the angle between the initial and the final electron momenta.

The next order terms of the cross section  are given by the interference between  $\mathcal{M}_{1\gamma}$ and  second order amplitude $\mathcal{M}^{(2)}$. In the complete calculation, in order to remove the infrared (IR) divergences, the inelastic Bremsstrahlung contribution must be also taken into account,
\begin{eqnarray}
d \sigma^{(2)}  & \sim &   2 \mathrm{Re}\left[ (i\mathcal{M}_{1\gamma})^* i\mathcal{M}^{(2)}\right] + d \sigma_{Brem.}^{(1)}.
\end{eqnarray}

In this paper we focus on the TPE box diagrams
(Fig. \ref{Fig_boxdiagrms}), which describe an exchange of  two photons between the electron and the proton target. The intermediate hadronic state is the off-shell nucleon or a resonance. Because the off-shell electromagnetic form factors are not known \cite{off_shell_vertex} we make a common ansatz and consider the on-shell vertices instead.

The leading  TPE contribution reads,
\begin{equation}
\label{I_TPE}
\mathcal{I}_{2\gamma} \equiv
  2\mathrm{Re}\left\{ (i\mathcal{M}_{1\gamma})^*
i\mathcal{M}_{2\gamma}
\right\}.
\end{equation}
The box diagrams contributing to $\mathcal{M}_{2\gamma}$ are drawn in Fig. \ref{Fig_boxdiagrms}.
\begin{figure}
\begin{center}
\includegraphics[width=0.5\textwidth]{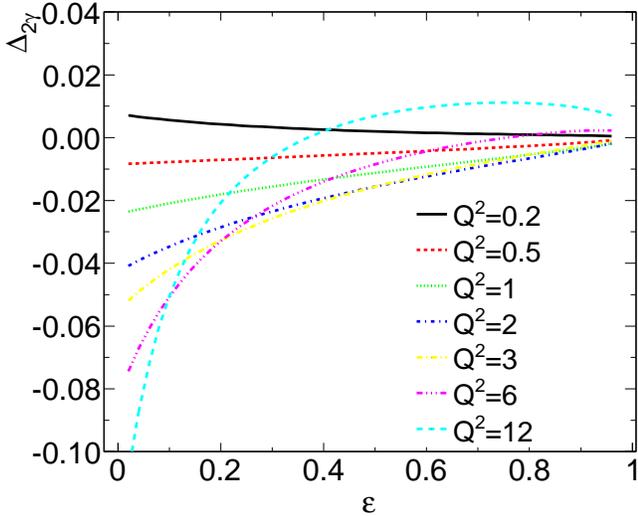}
\caption{(Color online) $\Delta_{2\gamma}$ (Eq. \ref{Delta_2gamma}) computed for  $\Box(N)$ TPE contribution.  The  form factors  from Ref. \cite{Blunden:2005ew} are used. The values of $Q^2$  are  in GeV$^2$ units.
\label{Fig_Blunden05}
}
\end{center}
\end{figure}

In the old $ep$  data analysis to account for higher order radiative corrections MT approach \cite{MT_radiative} was usually applied. In this approach the TPE box contribution was computed in the soft photon approximation. As it was pointed out by Blunden et al. \cite{Blunden:2003sp} to properly correct the cross section data by ''full'' TPE term,   one has to subtract first  the MT box contribution. Then the  redefined  TPE correction reads,
\begin{equation}
\label{Delta_2gamma}
\Delta_{2\gamma} = \delta_{2\gamma}(full) - \delta_{2\gamma}(MT),
\end{equation}
where
\begin{equation}
\delta_{2\gamma} = \frac{\mathcal{I}_{2\gamma}}{\displaystyle \left| i\mathcal{M}_{1\gamma}\right|^2},
\end{equation}
and  it is given by some integral, see Eq. 23 of Ref. \cite{Blunden:2005ew}.

The inclusion of the TPE correcting term modifies the form of the reduced cross section, namely,
\begin{equation}
\label{sigma_reduced_2photon}
\sigma_{R,1\gamma+2\gamma}(Q^2, \varepsilon) \to \sigma_{R,1\gamma}(Q^2, \varepsilon) + \Delta C_{2\gamma}(Q^2, \varepsilon),
\end{equation}
where $
\Delta C_{2\gamma}  = \Delta_{TPE}  \cdot \sigma_{R,1\gamma}.
$
\begin{figure}
\centering{
\includegraphics[width=0.5\textwidth]{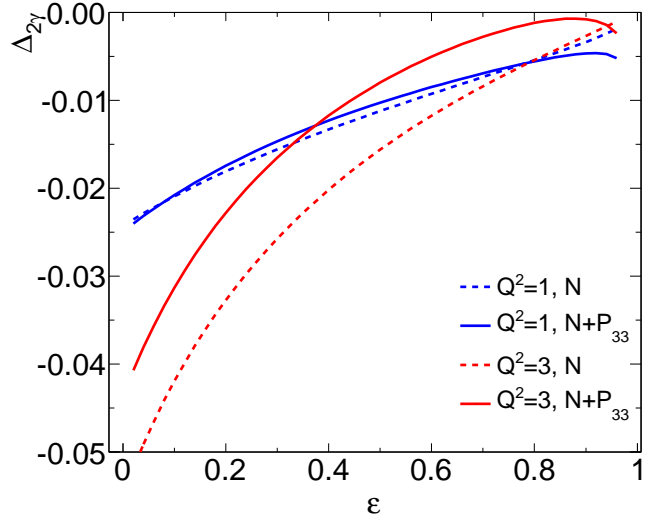}
\caption{(Color online) $\Delta_{2\gamma}$ (Eq. \ref{Delta_2gamma}) computed for either  $\Box(N)$ or $\Box(N+P_{33})$ TPE contributions. The  form factors  from Ref. \cite{Blunden:2005ew} are applied. The inelastic TPE correction, given by $\Box(P_{33})$,  is computed within  the  $P_{33}(full)$   model. The values of $Q^2$  are  in GeV$^2$ units.
\label{Fig_Blunden05ND}
}
}
\end{figure}

\section{Neural Network Approach}

\label{Section_NN_attempt}

The artificial neural networks (ANN) have been used in the particle and nuclear physics
for many years. The ANNs are perfectly  dedicated to particle or interaction identification and have been applied in the experimental data analyses  \cite{NN_in_Physics}. Study of the properties of  ANNs is an interesting topic   by itself. Neural networks have also been investigated within  the methods of the statistical physics \cite{Hertz_book}.

The feed-forward neural network is a type of the ANN,  which can be applied to  interpolating  data,   parameter estimation, and function approximation problems.

The ANN methodology can be a  powerful approach for doing the approximation of the physical observables based on the measurements   if it is difficult to make the predictions (base on the theoretical model) of the analysed  quantities  or  the theoretical predictions are  model-dependent but there exist the informative experimental data. Then one can construct a model-independent representation (given by neural network)  of the physical observables favoured by the measurements. As an example  let us mention the parton distributions functions (PDFs), which are  parametrized by the feed-forward neural networks \cite{Ball:2013lla}.

Similarly as in the case of the PDFs computing the nucleon FFs and the TPE correction from the first principles is  a difficult task.  On the other hand there are  unpolarized cross section, ratio $R_{+/-}$ as well as PT ratio  data distributed in the wide kinematical range.  Global analysis of these measurements  provide  reliable information about the  FFs and the TPE \cite{Arrington:2007ux,Graczyk:2011kh}.

The BNN approach was adapted by us \cite{Graczyk:2010gw,Graczyk:2011kh}  to approximate the nucleon FFs and TPE correction. In the next four subsections  a short review of the main features of this approach is presented.

\subsection{Multi-Layer Perceptron}
\label{Section_NN_attempt_MLP}

The FFs and TPE correction are going to be approximate by the feed-forward neural networks in the multi-layer perceptron (MLP) configuration. From the mathematical point of view the MLP, denoted as
$\mathcal{N}$, is a non-linear function, which maps
a subset of  $\mathbb{R}^{n_i}$ (an input space) into $\mathbb{R}^{n_o}$ (an output space) where $n_i,n_o \in \mathbb{N}$. Given MLP consists of several layers of units, namely, input,
hidden and  output layers (see Figs. \ref{simplenetwork} and \ref{Fig_network}).

A unit  is a single-valued real function called an activation function
$(f_{act.})$. For the argument it takes the weighted sum of outputs
from the  the previous layer  units,
\begin{equation}
f_{act.}\left(\sum_{i=0}^n w_i f_{act.}^i(previous\;layer)
\right),
\end{equation}
where $w_i$ ($i=1,2,...,n$) is the weight parameter.

An example of the typical unit is drawn in Fig. \ref{simplenetwork}. The weight parameters are established during the training i.e.  a process of finding the optimal weight configuration. In reality the optimal weight configuration  minimizes  some error function.

In our analysis the sigmoid,
\begin{equation}
\label{activation_function}
f_{act}(x) = \frac{1}{1 + \exp(-x)}
\end{equation}
is  taken  for the activation functions  in the hidden layer.  But in the case of the output units we consider the linear activation functions. The above choice is motivated by Cybenko theorem
\cite{Cybenko_Theorem}, which states that it is enough to consider
the MLP with one hidden layer, and sigmoid-like
functions there as well as the linear activation functions in the output layer to approximate any continuous function\footnote{According to the Cybenko theorem the discontinuous functions can be approximated well by the MLP with two hidden layers of units.}.

Notice that the effective support of (\ref{activation_function}) is  limited. It is a useful feature in the case of the numerical analysis (the weights are randomly initialized at the beginning of every training).

For a more detailed description of the MLP properties, the training process, learning algorithms etc. see Sect. 2 of Ref. \cite{Graczyk:2010gw}.

\subsection{Overfitting Problem}

\label{Section_NN_attempt_overfitting}

A simple example of the one-hidden layer MLP configuration,  used to approximate the electric proton FF \cite{Graczyk:2010gw}, is shown in Fig.~\ref{simplenetwork} (b). The input and output are the one-dimensional vectors, $(Q^2)$ and $(G_{E})$ respectively.
In this case the error function is postulated to be,
\begin{equation}
\label{Error_Function_FormFactors}
S_{ex}(\mathcal{D},\vec{w}) =\chi^2= \sum_{i=1}^N \left( \frac{G_{E}(Q^2_i;\vec{w})- G_{E,i}^{exp.}}{\Delta G_{E,i}^{exp.}} \right)^2.
\end{equation}
$N$ denotes the number of experimental points, $(G_{E,i}^{exp.},\Delta G_{E,i}^{exp.})$  is the $i$th experimental point (its best value and error). By $\mathcal{{D}}$ the experimental data set (or sets) are denoted.

It is obvious that increasing the number of units (degrees of freedom) improves the ability of the network for representing the data. A MLP with large enough  number of weight parameters  may fit to the data exactly but in this case the statistical fluctuation of the measurements are reproduced. Such  model has no predictive power and adding new data to the fit   spoils its quality. This kind of the network overfits the data (or it is said the network is overlearned). Such fit is characterized by unrealistic prediction of the uncertainties    (see discussion   in Sect. 2.1 and Figs. 3 and 4 of Ref. \cite{Graczyk:2010gw}).

One of the methods for facing the overfitting problem and finding the optimal network configuration  is to implement the Occam's razor principle. Then in a natural way   simpler network configurations are preferred. The simplest  idea is to consider a penalty term,
\begin{equation}
\label{Error_penalty}
\alpha E_w, \quad \textrm{where}\; E_w =  \frac{1}{2}\sum_{i=1}^W w_i^2.
\end{equation}
Including in the error function the above expression  may prevent from getting too large absolute  values of the weight parameters and as the result  the  overlearned networks\footnote{Usually the MLP that overfits the data contains at least one weight parameter  of  large absolute value.}.

The parameter $\alpha$ in (\ref{Error_Function_FormFactors}) is introduced to regularize the  penalty term. In general one would consider  several $\alpha$ parameters (each for every distinct class of weights), see e.g. Sect. 3.2 of Ref. \cite{Graczyk:2010gw} or Chapter 9 of Ref. \cite{Bishop_book}.

The major difficulty is to find an optimal value of the $\alpha$ parameter. The Bayesian framework offers mathematically consistent method for getting such $\alpha$'s. Indeed in this approach the  penalty term has a natural probabilistic interpretation and   $\alpha$  is computed within the objective Bayesian algorithm.

\subsection{Bayesian Neural Networks}

\label{Section_NN_attempt_BNN}

The bayesian framework  for the MLP \cite{Bishop_book,bayes} was developed to provide  consistent and objective methods, which allows one to:
\begin{itemize}
 \item establish optimal structure of the MLP (number of the hidden units, layers);

 \item find optimal values of the weights and the $\alpha$ parameters;

 \item establish  optimal values of the learning algorithm parameters;

 \item compute the neural network output uncertainty, and uncertainties for the weight and $\alpha$ parameters.
 \item classify  and compare    models  quantitatively.
\end{itemize}
The BNN approach requires  minimal  input from the user. Indeed the idea of the approach  was to replace the user's common sense by the mathematical objective procedures \cite{Bishop_book}. Obviously some user's input is necessary.

 Below we shortly review the  BNN approach. For more detailed description of the BNN  see Refs. \cite{Graczyk:2010gw} (Sect. 3), \cite{Graczyk:2011kh} (Sect. III) as well as \cite{Bishop_book} and  \cite{MacKay_thesis}.

\subsubsection{Model Comparison}

\label{Section_NN_attempt_Model_Comparison}

Let us consider a set  $\mathcal{S}$, which contains MLPs with different number of hidden units. Without loosing the generality of the approach we can restrict the set $S$ to the MLPs with only one hidden layer (the choice supported by Cybenko theorem). Each network $\mathcal{N}_\beta\in \mathcal{S} $ ($\beta=1,2,...$) approximates some physical quantities based on the data  $\mathcal{D}$. The models (networks) can be classified by a conditional probability
\begin{equation}
\label{P_Bayes}  P(\mathcal{N}_\beta|\mathcal{D}).
\end{equation}
The BNN approach gives a recipe how to construct and compute the above  function.

The Bayes' theorem connects the probability (\ref{P_Bayes}) with the so-called evidence $P(\mathcal{D}|\mathcal{N}_\beta)$,
\begin{equation}
\label{Probability_distribution_of_models}
P(\mathcal{N}_\beta|\mathcal{D}) = \frac{P(\mathcal{D}|\mathcal{N}_\beta)P(\mathcal{N}_\beta)}{P(\mathcal{D})}.
\end{equation}
$P(\mathcal{D})$ is the normalization factor, which does not depend on the model $\mathcal{N}_\beta$. $P(\mathcal{N}_\beta)$ is the prior probability. However, at the beginning of any analysis there is no reason to prefer a particular model (network) therefore  it is natural to assume that,
\begin{equation}
P(\mathcal{N}_1)=P(\mathcal{N}_2)=P(\mathcal{N}_3)=...
\end{equation}
Hence the evidence  differs from $P(\mathcal{N}_\beta|\mathcal{D})$   by only a constant normalization factor and  it can be used to qualitatively classify the statistical hypotheses.

We apply the so-called hierarchical approach \cite{MacKay_thesis} to construct and then to compute the evidence. It is several steps procedure, which  is described in the next four subsections.

\subsubsection{First Step}

 \label{Section_NN_attempt_First_Step}

 In the first  step of the approximation the  posterior probability distribution  $P(\vec{w}| \mathcal{D}, \alpha, \mathcal{N}_\beta)$ is computed. According to the Bayes' theorem it reads,
\begin{eqnarray}
\label{posterior_w}
P(\vec{w}| \mathcal{D}, \alpha,\{\mathcal{I}_{Phys.} \}, \mathcal{N}_\beta) &=&\nonumber\\
& & \!\!\!\!\!\!\!\!\!\!\!\!\!\!\!\!\!\!\!\!\!\!\!\!\!\!\!\!\!\!\!\!\!\!\!\!\!\!\!\!\!\!\!\!\!\!\!\!\!\!\!\!\!\!\!\!\!\!\!\!\!\!\!\!\!\!\!\!\!\!\!\!\!\!\!\!\!\!\!\!\!\!\!\!\!\!\!
\frac{\mathcal{P}\left(\mathcal{D}\right|\left.\vec{w}, \alpha,\{\mathcal{I}_{Phys.} \}, \mathcal{N}_\beta \right)
\mathcal{P}\left(\vec{w}\right|\left. \alpha,\{\mathcal{I}_{Phys.} \}, \mathcal{N}_\beta \right)}{\mathcal{P}\left(\mathcal{D}\right|\left. \alpha,\{\mathcal{I}_{Phys.} \}, \mathcal{N}_\beta \right)},
\end{eqnarray}
where
$\mathcal{P}\left(\vec{w}\right| \left.\alpha,\{I\}_{Phys.},  \mathcal{N}_\beta \right)$ is the prior probability, $\{\mathcal{I}\}_{Phys.}$ denotes the set of initial physical assumptions.

The likelihood function, $\mathcal{P}\left(\mathcal{D}\right|\left.\vec{w}, \alpha,\{\mathcal{I}_{Phys.} \}, \mathcal{N}_\beta \right) $ does not depend on $\alpha$ but in our analysis it is modified due to the physical constraints $\{\mathcal{I}\}_{Phys.}$,
\begin{eqnarray}
\mathcal{P}\left(\mathcal{D}\right|\left.\vec{w},  \mathcal{N}_\beta ,\{\mathcal{I}\}_{Phys.}\right)&=& \nonumber \\
& &
\!\!\!\!\!\!\!\!
\!\!\!\!\!\!\!\!
\!\!\!\!\!\!\!\!
\!\!\!\!\!\!\!\!
\!\!\!\!\!\!\!\!
\!\!\!\!\!\!\!\!
\!\!\!\!\!\!\!\!
\!\!\!\!\!\!\!\!
\frac{1}{n_\beta}
\exp\left(-S_{ex}(\mathcal{D},\vec{w}) - S_{Phys.}(\{\mathcal{I}_{Phys.}\},\vec{w})\right).
\end{eqnarray}
The normalization factor $n_\beta$ is computed in Hessian approximation, see Eq. (3.8) of Ref. \cite{Graczyk:2010gw}.

The  functions $S_{ex}(\mathcal{D},\vec{w})$ and $S_{Phys.}(\{\mathcal{I}_{Phys.}\},\vec{w})$ are given by  some $\chi^2$ distributions.  $S_{Phys.}(\{\mathcal{I}_{Phys.}\},\vec{w})$  is introduced to force the MLP to properly reproduce   the form factors at $Q^2=0$. (see Sec. III of Ref.\cite{Graczyk:2011kh}). It may also account  for other physical constraints.

The prior $\mathcal{P}\left(\vec{w}\right|\left.  \mathcal{N}_\beta \right)$ describes  only the initial ANN assumptions about the weights. A reasonable approximation is to  assume that it is given by the normal distribution, centred at $\vec{w}_0=\vec{0}$.
\begin{eqnarray}
\mathcal{P}\left(\vec{w}\right|\left. \alpha, \mathcal{N}_\beta \right)
&=& \frac{1}{n_{a}} \exp[- \alpha E_w] \\
n_{a} & = & \int d^{W_\beta} w \exp[- \alpha E_w].
\end{eqnarray}

The optimal configuration of weights $\vec{w}_{MP}$ maximizes the posterior probability (\ref{posterior_w}). In reality it minimizes the following error function,
\begin{equation}
\label{error_function}
S_{ex}(\mathcal{D},\vec{w}) + S_{Phys.}(\{\mathcal{I}_{Phys.}\},\vec{w}) + \alpha E_w.
\end{equation}

Notice that in this step of the approximation the $\alpha$ parameter is assumed to  be known.

The $1\sigma$ error of any physical observable $\mathcal{O}$, which depends on the network response is a square root of the variance,
\begin{eqnarray}
\left(\langle \Delta \mathcal{O}\rangle\right)^2 &=& \langle\mathcal{O}^2\rangle - \langle\mathcal{O}\rangle^2 \nonumber\\
\langle\mathcal{O}\rangle &  = &
\int d^{W_{\beta}} w \, \mathcal{O}(\mathcal{N}_\beta)  P(\vec{w}| \mathcal{D}, \alpha,\{\mathcal{I}_{Phys.} \}, \mathcal{N}_\beta). \nonumber \\
\end{eqnarray}
The above integral is computed in the Hessian approximation.

\subsubsection{Second Step}

\label{Section_NN_attempt_Second_Step}

The optimal value of the $\alpha$ parameter $(\alpha_{MP})$ maximizes the posterior probability,
\begin{eqnarray}
\label{posterior_alpha}
\mathcal{P}\left(\alpha\right|\left. \mathcal{D},\{\mathcal{I}_{Phys.}\}, \mathcal{N}_\beta \right)  &=& \nonumber \\
& &
\!\!\!\!\!\!\!\!\!\!\!\!\!\!\!\!\!\!\!\!
\!\!\!\!\!\!\!\!\!\!\!\!\!\!\!\!\!\!\!\!\!
\!\!\!\!\!\!\!\! \!\!\!\!\!\!\! \!\!\!\!\!\!  \frac{\mathcal{P}\left(\mathcal{D}\right|\left. \alpha,\{\mathcal{I}_{Phys.}\}, \mathcal{N}_\beta \right) \mathcal{P}\left(\alpha\right|\left. \{\mathcal{I}_{Phys.}\}, \mathcal{N}_\beta \right)}{\mathcal{P}\left(\mathcal{D}\right|\left.\{\mathcal{I}_{Phys.}\},\mathcal{N}_\beta \right)}.
\end{eqnarray}
(the denominator of the above expression  is obtained in the previous step of the approximation).

The necessary condition, which must by satisfied by $\alpha_{MP}$ reads,
\begin{equation}
\left.\frac{\partial }{\partial \alpha} \mathcal{P}\left(\alpha\right|\left. \mathcal{D}, \mathcal{N}_\beta \right)\right|_{\alpha=\alpha_{MP}} =0.
\end{equation}
It can be shown that in the Hessian approximation the above equation can be written as,
\begin{equation}
2 \alpha_{MP} E_{w}(\vec{w}_{MP}) =\sum_{i=1}^{W_\beta} \frac{\lambda_i}{\lambda_i + \alpha_{MP}} \equiv \gamma(\alpha_{MP}),
\end{equation}
where $\lambda_i$ are the eigenvalues of  $H_{kj}=\nabla_k \nabla_j (S_{ex}+S_{Phys.})$, $k,j=1,..., W_\beta$, $\nabla_k\equiv \partial/\partial w_k$.

In practice  $\lambda_i$'s depend  on $\alpha$. Hence to get the optimal $\alpha_{MP}$, the value of $\alpha$  is iteratively changed during the training, $\alpha_{k+1} = \gamma(\alpha_{k})/2 E_w$. It means that the optimal weights and $\alpha$ parameter are established during the same training process.

The initial value of the $\alpha$ parameter is taken to be large, which corresponds to the prior assumption that at the beginning of the analysis almost all relevant  values of weights    are   probable.

\subsubsection{Third Step}

\label{Section_NN_attempt_Third_Step}
In this step the evidence is computed. Notice that it is the denominator of right-hand side of Eq. \ref{posterior_alpha}. Careful calculations leads to the following expression for log of evidence (see Sect 3.1 of Ref. \cite{Graczyk:2010gw}),
\begin{eqnarray}
\ln  \mathcal{P}\left(\mathcal{D}\right|\left.\{\mathcal{I}_{Phys.}\},\mathcal{N}_\beta \right)
 & = & \nonumber
 \\
 & & \label{evidence_misfit}
 \!\!\!\!\!\! \!\!\!\!\!\! \!\!\!\!\!\! \!\!\!\!\!\! \!\!\!\!\!\!
 \!\!\!\!\!\! \!\!\!\!\!\! \!\!\!\!\!\! \!\!\!\!\!\! \!\!\!\!\!\!
 -S_{ex}(\mathcal{D},\vec{w}_{MP})
 -S_{Phys.}(\{\mathcal{I}_{Phys.}\},\vec{w}_{MP}) \\
& &
\!\!\!\!\!\! \!\!\!\!\!\! \!\!\!\!\!\! \!\!\!\!\!\! \!\!\!\!\!\!
\!\!\!\!\!\! \!\!\!\!\!\! \!\!\!\!\!\! \!\!\!\!\!\! \!\!\!\!\!\!\!\!\!
- \alpha_{MP} E_{w}(\vec{w}_{MP})
-\frac{1}{2}\ln \mathrm{det} A + \frac{W}{2}\ln \alpha_{MP} -\frac{1}{2}\ln \frac{\gamma}{2} \nonumber
\\
 \label{evidence_Occam_volume}
  \\
& &
\label{evidence_Occam_symmetr}
\!\!\!\!\!\! \!\!\!\!\!\! \!\!\!\!\!\! \!\!\!\!\!\! \!\!\!\!\!\!
\!\!\!\!\!\! \!\!\!\!\!\! \!\!\!\!\!\! \!\!\!\!\!\! \!\!\!\!\!\!\!\!\!
+ (g+t) \ln(2) + \ln(g!)+\ln(t!),
\end{eqnarray}
 $A =H(\vec{w}_{MP}) +  \alpha_{MP} I$.

Expression (\ref{evidence_misfit}) is the misfit of the approximated data. It is usually of the low-value. Terms (\ref{evidence_Occam_volume}-\ref{evidence_Occam_symmetr})
contribute to the Occam's factor. Indeed (\ref{evidence_Occam_volume}) takes  large values for the models, which overfit the data. In a typical MLP  some hidden units, in the given layer,  can be reordered without affecting the values of the network output. It means that for every MLP  there exist several  equivalent indistinguishable network configurations. It gives rise to the additional normalization factor (\ref{evidence_Occam_symmetr}), which must be included to properly define  the evidence. The symmetry factor presented above concerns the MLP used in the extraction of the TPE correction from the data (see Sect. \ref{Section_NN_attempt_Extraction_of_TPE} and Eq. \ref{network_gt}).

\subsubsection{General Scheme}

\label{Section_NN_attempt_General_Scheme}

Schematically the  approach discussed above can be  summarized as follows,
\begin{eqnarray}
 1th\; Step: & \to & P(\vec{w}| \mathcal{D}, \alpha,\{\mathcal{I}_{Phys.} \}, \mathcal{N}_\beta)\\
 2d\;Step: & \to & \mathcal{P}\left(\alpha\right|\left. \mathcal{D},\{\mathcal{I}_{Phys.} \}, \mathcal{N}_\beta \right) \\
 3d\;Step: & \to & \mathcal{P}\left(\mathcal{D}\right|\left.\{\mathcal{I}_{Phys.}\},\mathcal{N}_\beta \right)
\end{eqnarray}

We see that the evidence, and the other posterior probabilities may depend on physical assumptions. Obviously  their impact on the final results must be carefully discussed.

\subsection{Extraction of TPE}
\label{Section_NN_attempt_Extraction_of_TPE}

The formalism discussed above was applied to extract the proton FFs and TPE correction \cite{Graczyk:2011kh} from the world elastic $e^-p$ and $e^+ p$ scattering data. We utilized the unpolarized cross section, ratio $R_{+/-}$ and PT data. The first two types of observables depend on two input variables $Q^2$ and $\epsilon$, while the last one depends only on $Q^2$.

On the other hand the TPE correction is a function of two input variables, but the FFs depend on  $Q^2$ only. This property was  encoded  in the network configuration by  dividing MLP  into two sectors (see Fig. \ref{Fig_network}). In the first there are $g$ units  connected only with the $Q^2$ input, while in the other  there are $t$ units connected with both input units. We denote this network as,
 \begin{equation}
 \label{network_gt}
 \mathcal{N}_{g,t}\left(\pmatrix{Q^2 \cr \varepsilon}; \vec{w}\right) = \pmatrix{G_E^\mathcal{N} \cr G_M^\mathcal{N} \cr \Delta \tilde{C}_{2\gamma}^\mathcal{N}}.
 \end{equation}
\begin{figure}
\includegraphics[width=0.5\textwidth,height=80mm]{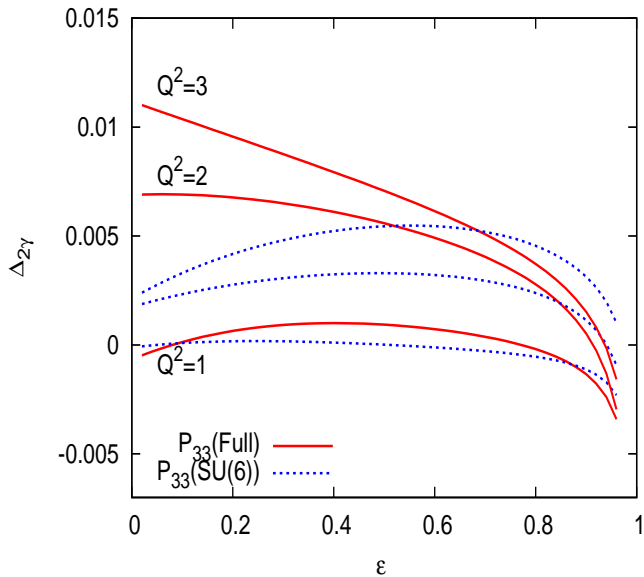}
\caption{(Color online) $\Delta_{2\gamma}$ (Eq. \ref{Delta_2gamma}) given by the resonance $P_{33}$ contribution only. Calculations are done for the $P_{33}(\mathrm{SU}(6))$  and  $P_{33}(full)$ models. The values of $Q^2$  are  in GeV$^2$ units. \label{Fig_delta_Blunden_vs_Full_SU6_multi_paper}}
\end{figure}

The BNN formalism seems to be well suited for performing a model-independent analysis but because the  utilized data turned out to be  not informative enough    some model assumptions had to be made.

The main constraint was induced by the following assumption,
\begin{enumerate}[A.]
\item
\label{assumption_ABGG_1}
The PT data is less sensitive to the TPE correction than the cross section measurements \cite{Guichon:2003qm}, hence the TPE contribution to $\mathcal{R}_{1\gamma}$ ratio can be neglected.
\end{enumerate}

As a consequence  the TPE correcting term was considered only in the case of the unpolarized cross section and $R_{+/-}$ data. Its extraction was induced by the presence of the PT measurements in the fit. Certainly it is an approximation,  therefore we  distinguish between $\Delta C_{2\gamma}$, as it is defined by theory, and $\Delta \tilde{C}_{2\gamma}$ as it is given by the BNN analysis. Both quantities enter  the reduced cross section formula  in the same way, see Eq. \ref{sigma_reduced_2photon}. But the latter is needed  to get a consistent fit of the CS,  $R_{+/-}$ and PT.

To get the FFs properly behaved at $Q^2=0$, and TPE term at $\epsilon=0$ (as it is suggested by the C-invariance \cite{Rekalo:2003km,Rekalo:2003xa,Rekalo:2004wa}), we introduced $S_{Phys.}$ (Eq. \ref{error_function}). It was  a $\chi^2$ function containing three artificial points (for details see Sect. III of Ref. \cite{Graczyk:2011kh}).

In order to find the optimal MLP configuration  45 different configurations\footnote{Number of units in the hidden layer, ($g$ and $t$) was varied.} of MLPs were trained. The largest evidence was obtained for the model $\mathcal{N}_{5,6}$.

In general the optimal fit should be given by an average (weighted by evidence) over all hypothetical models. In this case the physical observable  $\mathcal{F}$, which is a function of FFs and TPE, reads,
\begin{eqnarray}
\langle  \mathcal{F}(G_E,G_M, \Delta \tilde{C}_{2\gamma})
\rangle  &=& \nonumber\\
\label{partition_function_def}
& &
\!\!\!\!\!\!\!\!\!\!\!\!\!\!\!\!\!\!\!\!\!\!\!\!\!\!\!\!\!\!\!\!\!\!\!\int_{\mathcal{S}} D\mathcal{N}  \mathcal{F}(G_E^\mathcal{N},G_M^\mathcal{N}, \Delta \tilde{C}_{2\gamma}^\mathcal{N}) \mathcal{P}(\mathcal{N}|\mathcal{D}).
\end{eqnarray}
In reality the above  integral can be written as  discrete series,
\begin{eqnarray}
\label{discrete_series}
\left\langle  \mathcal{F}(G_E,G_M, \Delta
\tilde{C}_{2\gamma}) \right\rangle &=& \nonumber \\
& &
\!\!\!\!\!\!\!\!\!\!\!\!\!\!\!\!\!\!\!\!\!\!\!\!
\!\!\!\!\!\!\!\!\!\!\!\!\!\!\!\!\!\!\!\!\!\!\!\!
\!\!\!\!\!\!\!\!\!\!\!\!\!\!\!\!\!\!\!\!\!\!\!\!
\sum_{m=1}^{M}\sum_{g=1,t=1}^{g+t=m}
\mathcal{F}(G_E^{\mathcal{N}_{g,t}},G_M^{\mathcal{N}_{g,t}}, \Delta \tilde{C}_{2\gamma}^{\mathcal{N}_{g,t}}) \mathcal{P}(\mathcal{N}_{g,t}|\mathcal{D}),
\end{eqnarray}
where $M\in \mathbb{N}$.

It turned out that the  evidence for  $\mathcal{N}_{5,6}$ model was much larger then for the other  analysed configurations of networks. Hence the expression (\ref{discrete_series}) contains only one dominant term,
\begin{equation}
\left\langle  \mathcal{F}(G_E,G_M, \Delta
\tilde{C}_{2\gamma}) \right\rangle
\approx
\mathcal{F}(G_E^{\mathcal{N}_{5,6}},G_M^{\mathcal{N}_{5,6}}, \Delta \tilde{C}_{2\gamma}^{\mathcal{N}_{5,6}}).
\end{equation}

\section{ Hadronic Calculations}
\begin{figure*}
\centering{
\includegraphics[width=\textwidth]{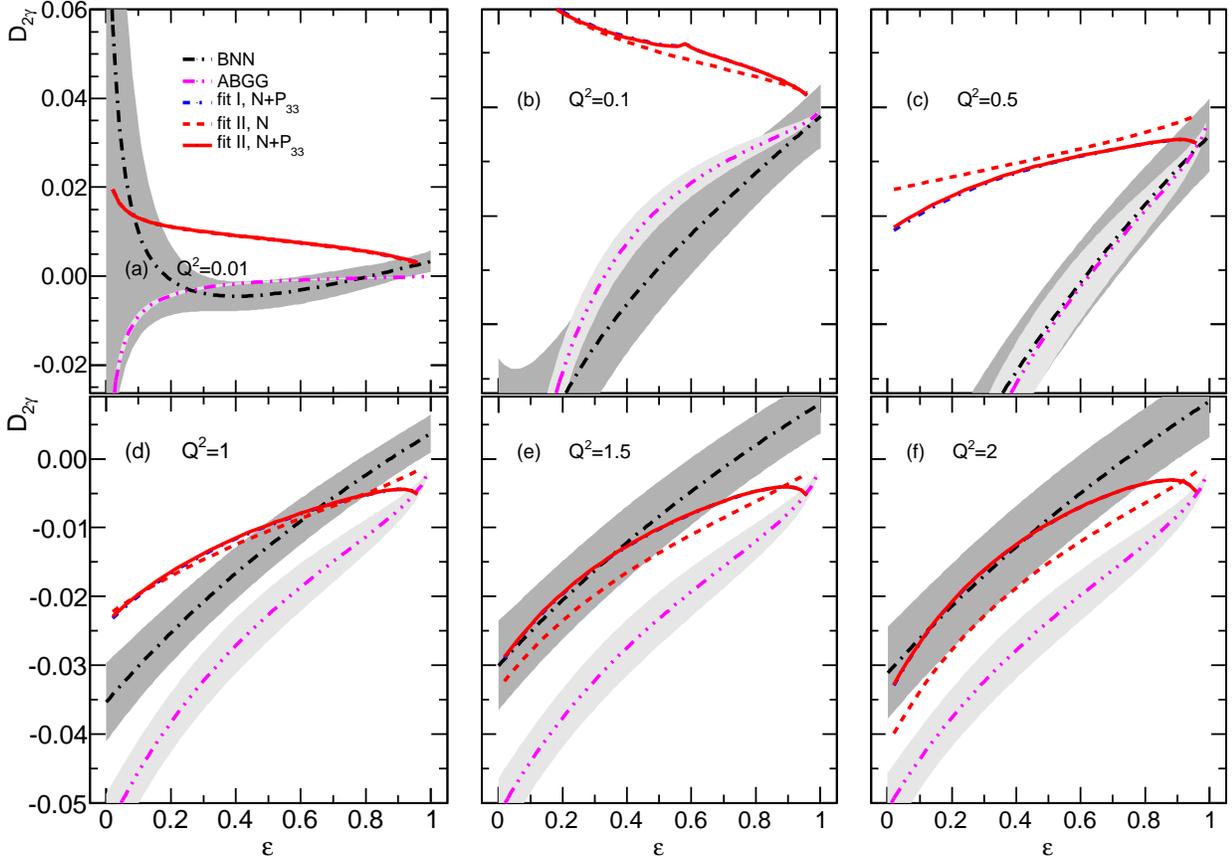}
\caption{(Color online) Predictions of $D_{2\gamma}$ (Eq. \ref{D_2gamma}) based on the BNN and HM (fits I and II) as well as ABGG  approaches. The TPE correction  includes either  elastic $(N)$ or  elastic and $P_{33}$ resonance (full model)  contributions. The values of $Q^2$  are  in GeV$^2$ units. The shaded areas show $1\sigma$ error computed from the covariance matrix.
\label{Fig_Delta2g_lowQ2_dependence}}
}
\end{figure*}
\begin{figure*}
\centering{
\includegraphics[width=\textwidth]{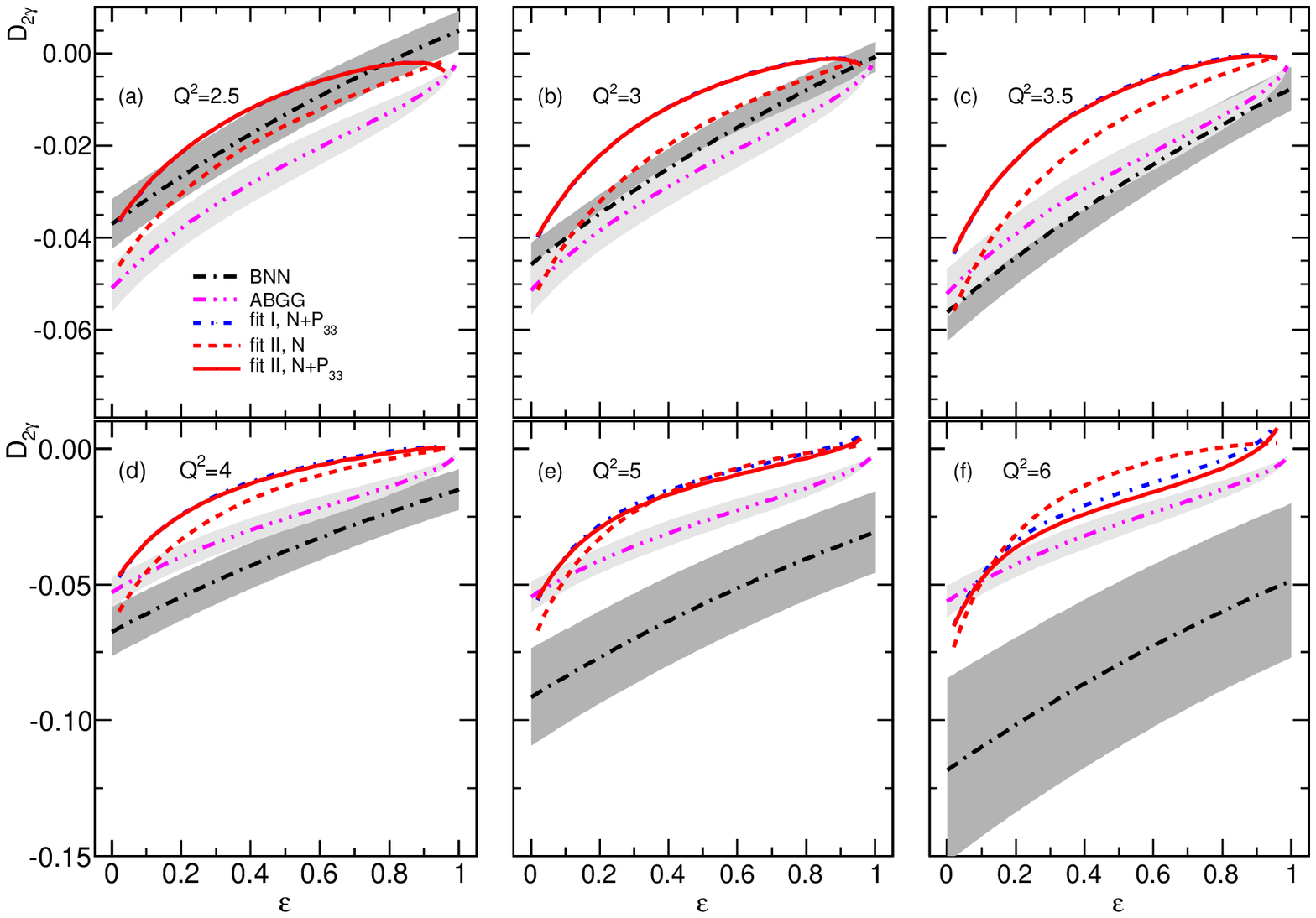}
\caption{(Color online) Caption the same  as in Fig. \ref{Fig_Delta2g_lowQ2_dependence}.
\label{Fig_Delta2g_highQ2_dependence}}
}
\end{figure*}

\label{Section_Theoretical_Attempt}

\subsection{Box Diagrams}

The TPE correction is computed in the similar way as in Refs. \cite{Maximon:2000hm,Blunden:2003sp,Blunden:2005ew,Kondratyuk:2005kk,Zhou:2009nf,Tjon:2009hf}.
Four box diagrams contribute to the $2\gamma$ amplitude (see Fig. \ref{Fig_boxdiagrms}): two with the nucleon intermediate hadronic state (denoted as $\Box(N)$) and two with $P_{33}(1232)$ hadronic intermediate state (denoted as $\Box(P_{33})$). The TPE contribution (\ref{I_TPE}) reads,
\begin{equation}
\label{TPE_correcting_term}
\mathcal{I}_{2\gamma} =
 2\frac{e^2}{Q^2}  \mathrm{Im}
 \left\{ w_{N}^\parallel +  w_{N}^\times + w_{\Delta}^\parallel + w_{\Delta}^\times  \right\},
\end{equation}
$w_{N,\Delta}^\parallel$ and  $w_{N,\Delta}^\times$ are the one-loop integrals represented by  direct and exchange  $\Box(N)$ and $\Box(P_{33})$ diagrams  respectively,
 \begin{eqnarray}
 \label{w_ND_paraller}
w_{N,\Delta}^{\parallel}  &=&
 \label{proton_box}
e^4 \int \frac{d^4 l}{(2\pi)^4}
\frac{\displaystyle L^{\alpha\mu\nu}_{\parallel} \mathcal{H}^{N,\Delta}_{\alpha\mu\nu} }{D(-k')}
\\
\label{w_ND_exchange}
 w_{N,\Delta}^{\times}
 &=&
 e^4 \int \frac{d^4 l}{(2\pi)^4}
 \frac{ \displaystyle L^{\alpha\mu\nu}_{\times} \mathcal{H}^{N,\Delta}_{\alpha\mu\nu} }{D(k)},
\end{eqnarray}
where
\begin{eqnarray}
D(x) & = & [(q+l)^2 + i\epsilon][ l^2 + i\epsilon] \times \nonumber \\
& & \!\!\!\!\!\!\!\!\!\!\!\!\!\!\!\!\!\!\!\!\! [(l+x)^2 - m^2 + i
\epsilon][(p'+l)^2 - M^2_{p,\Delta}+ i \epsilon].
\end{eqnarray}
We keep a nonzero electron mass $m=0.510$ MeV$/c^2$. $M_\Delta=1232$ MeV$/c^2$ denotes the $P_{33}$ resonance mass. The numerators of integrals  (\ref{w_ND_paraller}) and (\ref{w_ND_exchange})   are given by the contraction of three-dimensional leptonic with hadronic tensors.

The leptonic tensor is defined as follows,
\begin{equation}
L^{\alpha\mu\nu}_{\parallel,\times}
 \equiv
\sum_{spin} j^{\alpha*}j^{\mu\nu}_{\parallel,\times},
\end{equation}
where
\begin{eqnarray}
\label{box_N_direct}
j^{\mu\nu}_\parallel
& = &
\overline{u}(k') \gamma^\mu  (\hat{k'} - \hat{l}+m) \gamma^\nu u(k)
\\
\label{box_N_exchange}
j^{\mu\nu}_\times
& = &
\overline{u}(k') \gamma^\mu  (\hat{k} + \hat{l}+m) \gamma^\nu u(k),\\
\hat{x} & = & x_\mu \gamma^\mu.
	\end{eqnarray}
Four-vector $j^\alpha$ is given by Eq. (\ref{current_leptonic_OPE}).

We distinguish two types of the  hadronic tensors, one for the nucleon and another for the $P_{33}$ intermediate state,
\begin{eqnarray}
\mathcal{H}^{N,\Delta}_{\alpha\mu\nu} & \equiv & \sum_{spin} h_{\alpha}^* h_{\mu\nu}^{N,\Delta},
\end{eqnarray}
where $h_{\alpha}$ is given by Eq. (\ref{current_hadronic_OPE}) and
\begin{equation}
h_{\mu\nu}^N
=
\overline{u}(p') \Gamma_\nu(-l)  (\hat{p'} + \hat{l} + M_p)  \Gamma_\mu(q+l) u(p).
\end{equation}
The proton electromagnetic vertex $\Gamma_\mu$ is defined by Eq. \ref{1gamma_Nphoton_vertex}.
 The hadronic tensor for the $\Box(P_{33})$ diagrams  has the form,
\begin{eqnarray}
h_{\mu\nu}^\Delta
&=&
\overline{u}(p') \Gamma_{\mu\xi}^{\Delta,in}(-l,p'+l)  \left(\hat{p'} + \hat{l} + M_\Delta\right) \nonumber \\
& & \quad \Lambda^{\xi\eta}(p'+l)  \Gamma_{\nu\eta}^{\Delta,out}(q+l,p'+l) u(p).
\end{eqnarray}
$\Gamma_{\nu\mu}^{\Delta,out}(q_\Delta,P)$ and $\Gamma_{\nu\mu}^{\Delta,in}(q_\Delta,P)$  denote the vertex for the $\gamma^* N \to \Delta $ and $\Delta \to N \gamma^*$ transitions. For  more detailed definitions see Sec. \ref{Sect_P33_1232_Form_Factors}.

For the Rarita-Schwinger $3/2$ spin   field propagator we take,
\begin{equation}
S_{\mu\nu}^\Delta =- \frac{i(\hat{p} + M_\Delta)}{p^2 -
M_\Delta^2+i\Gamma_\Delta M_\Delta}\Lambda_{\mu\nu}(p).
\end{equation}
Similarly as  Kondratyuk et al. \cite{Kondratyuk:2005kk} we set
$\Gamma_\Delta \to 0$\footnote{Recently Borisyuk and Kobushkin \cite{Borisyuk:2013hja} performed calculations in which the impact of the nonzero value of  $\Gamma_\Delta$ is discussed.}. With this simplification the dominant contribution to the loop integrals comes from the $P_{33}$ resonance mass pole. Hence the choice of on-shell projection operator,
\begin{equation}
\label{Projection_33}
\Lambda_{\mu\nu}(p)= g_{\mu\nu} - \frac{1}{3}\gamma_\mu
\gamma_\nu -\frac{2p_\mu p_\nu}{3 M_\Delta^2} + \frac{p_\mu
\gamma_\nu - p_\nu \gamma_\mu}{3 M_\Delta} ,
\end{equation}
leads to the same results as  the off-shell projection operator discussed by Kondratyuk et al..  Taking into consideration this  approximation for projector operator simplifies and also accelerates the algebraic decomposition of the integrals (\ref{w_ND_paraller}) and (\ref{w_ND_exchange}). The procedure of computing the integrals (\ref{w_ND_paraller}-\ref{w_ND_exchange}) is described in the Appendix \ref{Appendix_A}.

\subsection{ Nucleon Form Factors}

For the nucleon FFs we consider two parametrizations:
\begin{itemize}
\item  parametrization I, sum of three monopoles,
\begin{equation}
\label{FF_polesum}
F_{k}(Q^2) =
\sum_{i=1}^3 \frac{f_i^k}{ m_i^k + Q^2 },
\end{equation}
where $f^1_3 = m^1_3 ( 1 -   f^1_1/m^1_1 - f^1_2/m^1_2)$,
$f^2_3 = m^2_3 ( \kappa_p  -   f^2_1/m^2_1 - f^2_2/m^2_2)$;

\item  parametrization II, sum of three dipoles,
\begin{equation}
\label{FF_dipolesum}
F_{k}(Q^2) = \sum_{i=1}^3 {f_i^k}\left( 1 + \frac{Q^2}{(m_i^k)^2}\right)^{-2},
\nonumber
\end{equation}
where $f_3^1 =  1 - f_1^1 - f_2^1$, $f_3^2 =  \kappa_p - f_1^2 - f_2^2$.
\end{itemize}

The parametrization I was previously discussed by  Blunden, et al. \cite{Blunden:2005ew} (BMT05).  In order to cross check  our algebraical and numerical procedures we repeat and check the calculations done in \cite{Blunden:2005ew}. In Fig. \ref{Fig_Blunden05} we present predictions  for $\Delta_{2\gamma}$ obtained for the same kinematics and the form factors as in BMT05 paper (for comparison see the plots in Figs. 2 and 3a of Ref. \cite{Blunden:2005ew}). We notice the excellent agreement between our hadronic model  and the BMT05 predictions.

 The $\Delta_{2\gamma}$ (or $D_{2\gamma}$) function depends weakly  on the proton form factors parametrization. Small differences between TPE predictions based on the parametrizations I and II  appear for larger values of $Q^2$  (see Figs. \ref{Fig_Delta2g_lowQ2_dependence} and \ref{Fig_Delta2g_highQ2_dependence}). But this is the region where the validity of the theoretical approach can questionable.

\subsection{  $P_{33}(1232)$ Form Factors}

\label{Sect_P33_1232_Form_Factors}

The hadronic vertex, $\Gamma_{\mu\nu}^{\Delta,out}(q,P_\Delta)$ for $\gamma^* p \to
\Delta^{++}$ transition is obtained by assuming that $P_{33}$ resonance is described by the Rarita-Schwinger 3/2 spin field,
\begin{equation}
\label{P33_vertex}
\overline{\Psi}^\nu(P)
\Gamma_{\mu\nu}^{\Delta,out}(q,P)  u(p), \quad q= P -p.
\end{equation}
One of the commonly discussed vertex parametrizations  is the following \cite{Jones:1972ky},
\begin{eqnarray}
\Gamma_{\mu\nu}^{\Delta,out}(q \equiv P-q,P) &=& \left[
\frac{C_5^V}{M^2} (g_{\mu\nu} p \cdot q -  {p}_\mu{q}_\nu) \right. \nonumber \\
& &
\!\!\!\!\!\!\!\!\!\!\!\!\!\!\!\!\!\!\!\!\!\!
\!\!\!\!\!\!\!\!\!\!\!\!\!\!\!\!\!\!\!\!\!\!
\!\!\!\!\!\!\!\!\!\!\!\!\!\!\!\!\!\!\!\!\!\!
\left.
      + \frac{C_4^V}{M^2}( g_{\mu\nu} q\cdot {P} -  {P}_\mu q_\nu) +\frac{C_3^V}{M}(g_{\mu\nu} \hat{q} -\gamma_\mu{q}_\nu)\right]\gamma_5. \nonumber\\
\end{eqnarray}
$P$ is the four-momentum of the outgoing $P_{33}$ resonance, while $p$ denotes the four-momentum of the incoming proton.

The $\Delta^{++} \to \gamma^* p$ vertex reads \cite{Kondratyuk:2005kk},
\begin{eqnarray}
\Gamma_{\mu\nu}^{\Delta,in}(p,P_\Delta) = \gamma_0
\left(\Gamma_{\mu\nu}^{\Delta,out}(p,P_\Delta) \right)^\dagger
\gamma_0.
\end{eqnarray}

For the $N\to P_{33}$ transition  form factors we consider two  scenarios,
\begin{itemize}
\item $P_{33}({\mathrm{SU}(6)})$ model: there is only
one vector form factor $C_3^V$; two other are obtained assuming
$\mathrm{SU}(6)$ quark model relations \cite{Liu:1995bu}, namely,
\begin{equation}
\label{su6_cv_limit} C_5^V(Q^2) = 0, \quad C_4^V(Q^2) = -
\frac{M}{M_\Delta}C_3^V(Q^2).
\end{equation}
In this case we parametrize the $C_3^V(Q^2)$ form factor as follows \cite{AlvarezRuso:1998hi},
\begin{equation}
\label{cv_old2} C_3^V(Q^2) = \frac{2.05}{\displaystyle \left(1  +
\frac{Q^2}{0.54 \, \mathrm{GeV}^2}\right)^2}.
\end{equation}
\item $P_{33}({Full})$ model: we apply the form factors from Ref. \cite{Lalakulich:2006sw}, namely,
\begin{equation}
\label{full_P33_form_factors}
C_i^V(Q^2) =   c_i^V \left( 1 +\frac{Q^2}{a_i
M_V^2}\right)^{-1}G_D(Q^2),
\end{equation}
where $a_3=a_4=4$, $a_5=0.776$, $c_3^V=2.13$, $c_4^V=-1.51$, $c_5^V=0.48$,
\begin{equation}
G_D(Q^2) = \left(1+\frac{Q^2}{M_V^2} \right)^{-2}, \quad
\mathrm{and}\quad M_V =0.84~\mathrm{GeV}.
\end{equation}
\end{itemize}

In the wide $Q^2$ range (for $Q^2 > 0.1$ GeV$^2$ the form factors given by Eq. (\ref{full_P33_form_factors})  take  the similar values as  MAID07 form factors \cite{Drechsel:2007if}.

The $P_{33}(full)$ model is different than the one  applied by Kondratyuk et al. \cite{Kondratyuk:2005kk} (denoted as KBMT05). However, in the intermediate $Q^2$ range the predictions are comparable, as seen in comparing  our Fig. \ref{Fig_Blunden05ND} with Fig. 2  from Ref. \cite{Kondratyuk:2005kk}). We notice also a qualitative agreement   with predictions  Borisyuk and Kobushkin model  \cite{Borisyuk:2012he}. In both approaches the TPE $\Box(P_{33})$ correction is positive at low and intermediate $Q^2$ range and it reduces the total TPE correction.

In contrast to $\Delta_{2\gamma}(N)$, the function $\Delta_{2\gamma}(P_{33})$  depends on the details of the hadronic model. Indeed, there are small but noticeable differences between predictions KBMT05 and $P_{33}(full)$ model.
To  illustrate the model dependence of $\Delta_{2\gamma}(P_{33})$  the prediction of TPE correction obtained within $P_{33}(\mathrm{SU}(6))$ and $P_{33}(full)$ models are plotted in Fig. \ref{Fig_delta_Blunden_vs_Full_SU6_multi_paper}. There is a clear discrepancy between predictions of both approaches.

\begin{figure}
\centering{ 
\includegraphics[width=0.45\textwidth]{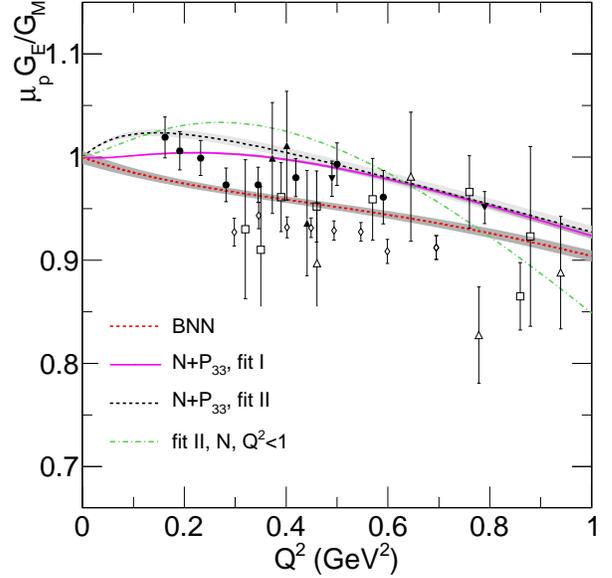}
\caption{(Color online)
Ratio $\mu_p G_E/G_M$ calculated based on fits I and II ($N$ and resonance $P_{33}$ contributions)  as well as BNN fit. Additionally ratio $\mu_p G_E/G_M$ obtained based on  the  fit (parametrization II) to the unpolarized cross section data below $Q^2=1$ GeV$^2$ is also plotted. The PT data (points with error bars) are taken from Ref. \cite{PT_data} and Zhan et al. \cite{Zhan:2011ji} (open diamonds). The shaded areas show $1\sigma$ error computed from the covariance matrix.
\label{Fig_ratio_ff_small}
}
}
\end{figure}

\begin{figure*}
\centering{
\includegraphics[width=\textwidth]{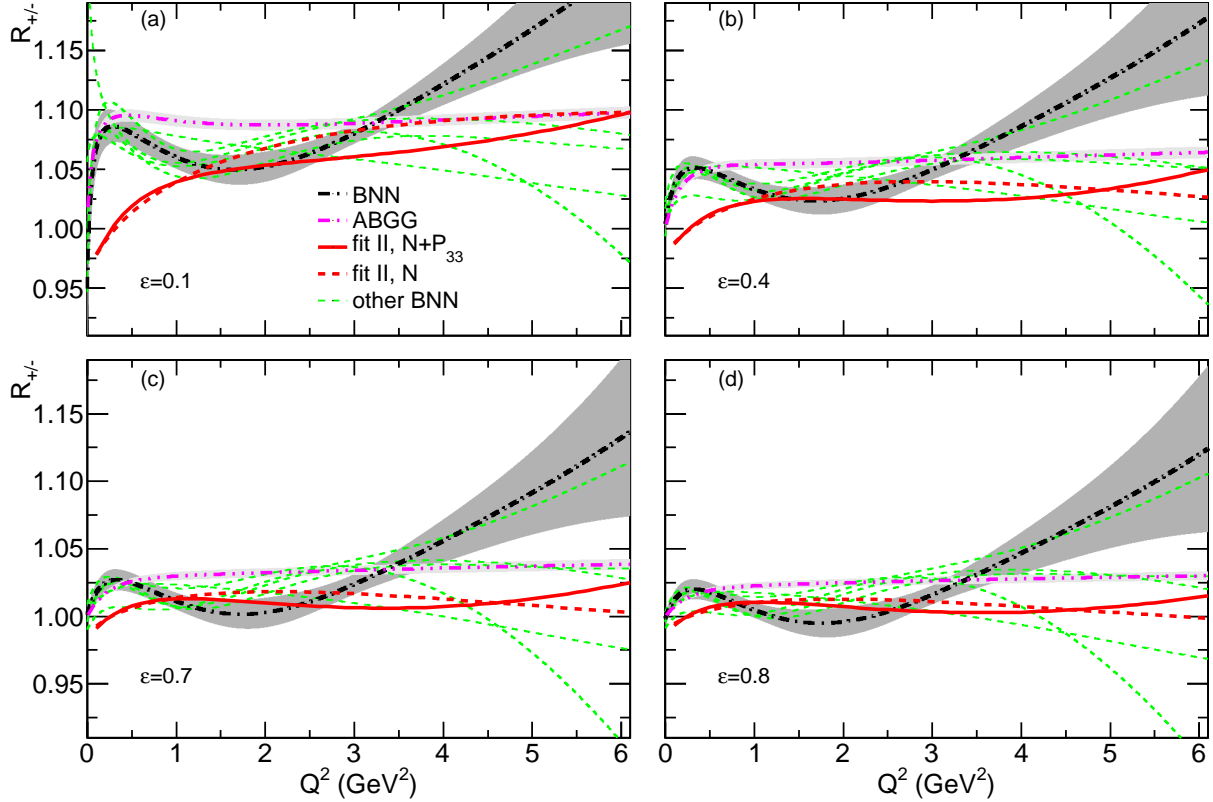}
\caption{(Color online) Predictions of  $R_{+/-}$ (Eq. \ref{R_+_-}) computed based on the BNN, HM (fit II) and ABGG approaches.  The TPE correction  includes either  elastic $(N)$ or  elastic and $P_{33}$ resonance (full model)  contributions.   Additionally the plots of $R_{+/-}$  predicted based on  the BNN fits rejected due too small value of the evidence (see Tab. \ref{Tab_chi2_min}), are presented (other BNN). The shaded areas show $1\sigma$ error computed from the covariance matrix.
\label{Fig_Ratio_pos_eps_dependence}}
}
\end{figure*}

\section{Neural Network vs. Hadronic Model}
\begin{figure}
\centering{
\includegraphics[width=0.5\textwidth]{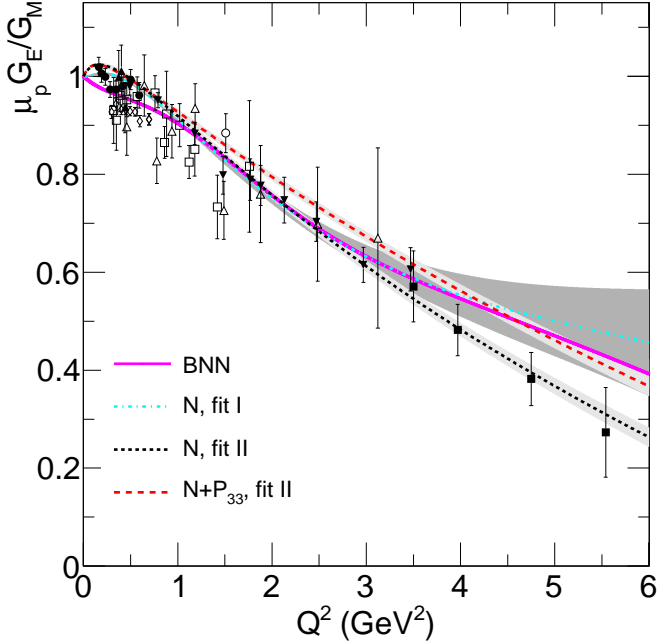}
\caption{(Color online)
Ratio $\mu_p G_E/G_M$ computed based on the fits I and II as well as for BNN. The HM fits include either elastic $(N)$ or elastic and $P_{33}$ resonance contributions. The PT data are taken from Refs. \cite{PT_data,Zhan:2011ji}. The shaded areas show $1\sigma$ error computed from the covariance matrix.
\label{Fig_ratio_ff_N_vs_ND}
}
}
\end{figure}

\label{Section_Results}


\begin{table}
\begin{tabular}{c|r|r}
\hline
 & $k=1$ & $k=2$  \cr
\hline
 &       &        \cr
$m_1^k$ &   $ 1.234 $     &    $ 0.321  $    \cr
$m_2^k$ &   $  0.181 $     &    $ 4.298 $    \cr
$m_3^k$ &   $ 1.085 $     &    $ 4.641 $     \cr
$f_1^k$ &   $ -6.569  $     &    $ 0.694  $  \cr
$f_2^k$ &   $ 0.055  $     &    $ -13.44 $   \cr
 \end{tabular}
 \quad
 \begin{tabular}{c|r|r}
 \hline
  & $k=1$ & $k=2$  \cr
 \hline
  &       &        \cr
 $m_1^k$ &   $ 1.221 $     &    $ 0.327$    \cr
 $m_2^k$ &   $ 0.173 $     &    $ 4.019 $   \cr
 $m_3^k$ &   $ 1.097 $     &    $ 4.450 $    \cr
 $f_1^k$ &   $  -7.934 $     &    $ 0.713 $   \cr
 $f_2^k$ &   $  0.051 $     &    $ -10.16 $   \cr
  \end{tabular}
\caption{\label{Tab_polesumN} Form factor parameters for the fit I  (\ref{FF_polesum})  for the hadronic model with elastic (N) (left panel) and  elastic and resonance $P_{33}$ (right panel) TPE contributions.  Mass parameters are in the units of $GeV$.}
\end{table}

The electromagnetic FFs of the proton  are the input of the hadronic model used in this paper.  For the comparison self-consistency  the proton FFs of HM are obtained from the fit of HM to the same unpolarized cross section data as in the BNN (for details see Appendix \ref{Appendix_B}).   The PT and $R_{+/-}$ data are not taken into consideration, and the constraint  coming from the assumption (\ref{assumption_ABGG_1}) does not affect the results. The TPE correction contains either $\Box(N)$ or  $\Box(N+P_{33})$ contributions. The obtained FFs parameters  are given in Tables. \ref{Tab_polesumN} (fit I) and \ref{Tab_dipolesumN} (fit II), while the values of $\chi^2_{min}/NDF$ (NDF = number degrees of freedom) are reported in the table below.\\

\begin{center}
\begin{tabular}{c|c|c}
\hline
FF  & $(N)$    &  $(N+P_{33})$ \\
\hline
fit I & $389/403$ &  $397/403$ \\
& & \\
fit II & $386/403$ &  $395/403$ \\
\end{tabular}
\end{center}

It is interesting to notice that the mass parameters of the fit I are not well-spaced. For instance the parameters $m_1^1$ and $m_3^1$ take quite similar values. The same feature characterizes the fits from Ref. \cite{Blunden:2005ew}, where the parametrization I was also discussed but it was fitted to the FFs from Ref. \cite{Mergell:1995bf}.  Indeed, this parametrization at large $Q^2$ behaves as $1/Q^2$, while it is expected (based on the theoretical arguments \cite{Kelly:2004hm,Brodsky:1973kr}) that  $G_{E,M} \sim 1/Q^4$.

The above observations may suggest that  the parametrization I is too simple to  describe accurately the FFs in the wide $Q^2$ range. In order to verify this statement we make two fits. In the first we consider the  data  below  $Q^2=$ 1 GeV$^2$, while in the other  we use the data below $Q^2=0.5$ GeV$^2$. For the first case we get the mass parameters $m_1^1=2.59$, $m_2^1=0.95$ and $m_3^1=0.18$ and for the other  $m_1^1=1.70$, $m_2^1=0.21$ and $m_3^1=9.16$.  We see that for low $Q^2$-data fits the mass parameters are well separated. However, because the problems remarked above in further discussion the HM with the FFs given by fit I is  treated as a toy model, discussed to present  the systematic properties of the hadronic approach.
 \begin{table}
\begin{tabular}{c|c|c}
\hline
 & $k=1$ & $k=2$ \cr
\hline
 &       &       \cr
$m_1^k$ &   $ 0.7732 $     &    $ 1.0595 $     \cr
$m_2^k$ &   $ 0.9489 $     &    $ 1.5629 $     \cr
$m_3^k$ &   $ 0.8457 $     &    $ 0.5474 $     \cr
$f_1^k$ &   $ 3.9833 $     &    $ 1.2645 $     \cr
$f_2^k$ &   $ 3.9334 $     &    $ -0.269 $     \cr
 \end{tabular}
 \quad
 \begin{tabular}{c|c|c}
 \hline
  & $k=1$ & $k=2$ \cr
 \hline
  &       &       \cr
 $m_1^k$ &   $ 0.7866 $     &    $  1.0247 $     \cr
 $m_2^k$ &   $ 0.9641 $     &    $  1.4914 $     \cr
 $m_3^k$ &   $ 0.8550 $     &    $  0.5082 $     \cr
 $f_1^k$ &   $ 4.3360 $     &    $  1.4592 $     \cr
 $f_2^k$ &   $ 3.7328 $     &    $  -0.3175 $    \cr
  \end{tabular}
\caption{\label{Tab_dipolesumN} Form factor parameters for the fit II  (\ref{FF_dipolesum})  for the hadronic model with elastic (N) (left panel) and  elastic and resonance $P_{33}$ (right panel) TPE contributions. Mass parameters are in the units of GeV. }
\end{table}

At low-$Q^2$ there is a visible discrepancy between the BNN and the hadronic model FFs fits. It is illustrated in Fig. \ref{Fig_ratio_ff_small}, where the ratio  $\mu_p G_E/G_M$ is plotted. There is a satisfactory agreement between the fits I and II.  On the contrary the ratio $\mu_p G_E/G_M$ predicted by the BNN approach is more  consistent with the recent PT measurements  \cite{Zhan:2011ji} (these data  were not included in the BNN fit).

\begin{figure}
\centering{
\includegraphics[width=0.5\textwidth]{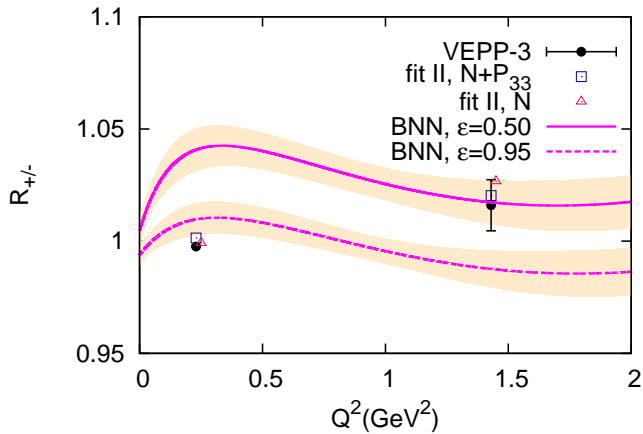}
\caption{(Color online) The ratio $R_{+/-}$  computed based on the BNN (lines) and HM (squares and triangles) approaches as well as the preliminary   VEPP-3 measurements \cite{Gramolin:2011tr} (filled circles). The HM predictions are computed  for the model (fit II) which contains  either elastic $(N)$ or elastic and resonance $P_{33}$ TPE contributions. The triangle points are  right shifted by 0.02 GeV$^2$.  The shaded areas show $1\sigma$ error computed from the covariance matrix.
\label{Fig_positron_data_vs_loops}}
}
\end{figure}

The low-$Q^2$ discrepancy between HM and BNN approaches is the result of different treatment  of the TPE corrections. It is illustrated in Fig. \ref{Fig_Delta2g_lowQ2_dependence} where we plot the function,
\begin{equation}
\label{D_2gamma}
D_{2\gamma} =
\frac{\Delta_{2\gamma}}{1 + \Delta_{2\gamma}}  = \frac{\Delta C_{2\gamma}}{\sigma_{R,1\gamma+2\gamma}}.
\end{equation}
 It can be  seen that the BNN and HM predictions are inconsistent for $Q^2 \in (0.02, 1)$ GeV$^2$. On the  contrary, below $Q^2 < 0.02$ GeV$^2$ and at low $\varepsilon$  there is a good agreement between  TPE predictions  obtained within  both methodologies  as well as  other theoretical calculations \cite{Coulomb_distortion}. This low $\varepsilon$ and $Q^2$ behaviour of the BNN fit seems to be a systematic property of all BNN-based parametrizations. It is illustrated in Fig. \ref{Fig_positron_data_vs_loops}, where the $R_{+/-}$ predicted by the BNN models, rejected due to too small value of the evidence (see Table \ref{Tab_chi2_min}), are plotted.  In the limit of $\varepsilon \to 0$, with $Q^2$ very low but fixed,   $\sigma_{R,1\gamma}$ (Eq. \ref{sigma_reduced}) is dominated by  the magnetic contribution, and the main constraint comes from the fact that  $G_M(Q^2=0)=\mu_p$.  As the result the TPE fit is affected by several data points  present at  low $Q^2$ and $\varepsilon$ domain.

For completeness of the   low-$Q^2$ comparison  we report  the values of the proton radius obtained from the BNN and HM fits. \\

\begin{tabular}{c|c|c|c}
          & BNN & fit I & fit II \cr
\hline
$\sqrt{\langle r^2_E \rangle}$ (fm) & $0.85 \pm 0.01$ & $0.898 \pm 0.001$ & $0.867\pm 0.002$\cr
          & & &
\end{tabular}\\
The value of  $\sqrt{\langle r^2_E \rangle}$ computed from  the BNN fit  is consistent  with the fit II and the recent atomic measurement  \cite{Pohl:2010zza} (0.84087(39) fm).  However, it disagrees with the prediction based on the fit I. The latter is inconsistent with the fit II as well.

There  are two major reasons  for the above inconsistency. The first one is induced by the systematic differences between the predictions of the TPE by BNN and HM approaches in the low $Q^2$ range.   The discrepancy between the values of the proton radius based on the fits I and II  is the result of the problems  of parametrization I (mentioned already above) with the proper description  of the FFs in the wide $Q^2$ range. In general the low number of parameters in  fits I and II  limits the flexibility of the FF parametrizations and their ability  for simultaneous description of the low and high $Q^2$ data. Therefore the low/high $Q^2$ fit dependence can be affected by the high/low $Q^2$ data.

To summarize, the low-$Q^2$ discussion we would like to emphasize that in both the present and the BNN data analyses our attention was not particularly focused  on the $Q^2\to 0$ limit. Certainly, accurate  calculations of the proton radius  require more careful  discussion, as it is reported in Refs. \cite{Arrington:2006hm,Zhan:2011ji,Bernauer:2010wm,Arrington:2011kv,Arrington:2012dq,Pohl:2013yb}.

Above $Q^2 = 1$ GeV$^2$  the BNN FFs ratio $\mu_p G_E/G_M$,   on the qualitative level, is comparable with the hadronic model predictions (Fig. \ref{Fig_ratio_ff_N_vs_ND}). All fits agree well with the PT  measurements \cite{PT_data}. As it could be expected the inclusion into the hadronic model of the $\Box(P_{33})$ contribution increases the value of the electric form factor at larger values of $Q^2$ (see Fig. \ref{Fig_ratio_ff_N_vs_ND}).

 An excellent consistency between predictions of the TPE effect by the BNN and HM approaches  appears for $Q^2\in (1,3)$ GeV$^2$ (Figs. \ref{Fig_Ratio_pos_eps_dependence} as well as \ref{Fig_Delta2g_lowQ2_dependence} and \ref{Fig_Delta2g_highQ2_dependence}). Above $Q^2 = 3$ GeV$^2$ the agreement is on the $2\sigma$ level only.

 In order to show the strength of the BNN approach we confront  its  predictions of the TPE  effect  with our previous global analysis (ABGG) \cite{Alberico:2008sz} made in the conventional way (see Figs. \ref{Fig_Delta2g_lowQ2_dependence} and \ref{Fig_Delta2g_highQ2_dependence}). In this approach,  following the proposal of Ref. \cite{Chen:2007ac}, some  functional form of the TPE term was postulated. But the same cross section and the PT data, as in the case of the BNN  were analysed. The constraint (\ref{assumption_ABGG_1}) was also imposed.  Although   both the electric and the magnetic FF  fits of ABGG analysis are very similar to those obtained within the BNN and the other phenomenological approaches \cite{Qattan:2011ke}, the predictions of the TPE correction  agree with the HM  only for $Q^2$ around 3 GeV$^2$ (see Fig. \ref{Fig_Ratio_pos_eps_dependence}). In the ABGG approach the model-dependence of the final fits was not discussed. The successful fits were characterized by reasonable value of $\chi^2_{min}/NDF$. However,  in the BNN analysis the models rejected, due too  low  evidence, were characterized also by the reasonable $\chi^2_{min}/NDF$ (see Table \ref{Tab_chi2_min}). But the TPE corrections predicted based on these fits, similarly as for ABGG are  inconsistent with the best BNN fit and the hadronic model calculations.
 \begin{table}[h]
 \centering{
 \begin{tabular}{c|c|c|c|c|c|c|c}
 \hline
 & $\mathcal{N}_{4,2}$  & $\mathcal{N}_{4,3}$ & $\mathcal{N}_{6,2}$ & $\mathcal{N}_{6,3}$ & $\mathcal{N}_{6,4}$  &  $\mathcal{N}_{5,7}$ & $\mathcal{N}_{5,6} $\\
 \hline
 $\chi^2_{min}$ & $ 507 $         &  $ 511 $       & $ 497 $        & $ 493 $        & $ 486 $         & $ 539 $          & $ \mathbf{478} $ \\
 ln(evidence) & $-633$  &  $-630$ & $-635$ & $-624$ & $-639$ &  $-699$ & $\mathbf{-611}$
 \end{tabular}
 \caption{The minimum of $\chi^2$ and the maximum of the evidence obtained for the best BNN model (bold fonts) and the fits rejected because of too small value of the evidence. A total number of points in the fit is 529.\label{Tab_chi2_min}}
 }
 \end{table}

The results of this paper are complementary to the conclusions
of Ref. \cite{Arrington:2007ux} (AMT). In this paper the global analysis of the world $ep$ data was performed.  The TPE correction was  given by a sum of the elastic  $\Box(N)$ and inelastic contributions. The latter was described by a phenomenological function, which approximates the resonance \cite{Kondratyuk:2006ig} and GPD-based \cite{Afanasev:2005mp}  fraction of the TPE effect. To compute the elastic contribution the FFs (parametrization I) were fitted to the electromagnetic FFs  from  \cite{Mergell:1995bf}. The FFs (parametrization from  \cite{Kelly:2004hm}) were fitted to the unpolarized cross section data (corrected by the TPE) and the PT measurements.  It was  shown that the cross section data modified due to the TPE effect are consistent with the PT measurements.  An effort was made to estimate uncertainties of the theoretical model for the TPE effect.

In  the AMT  as well as in the ABGG approaches the analyses were performed in the spirit of
frequentis  statistics (using the least square method), while the neural network analysis was done within the Bayesian statistics (for the short review see \cite{PDG}). In both statistical approaches to find the best fit some error function is minimized. But in the BNN approach the procedure of finding the optimal model is more complicated. In the first stage of the approach the large population  of the MLPs (more than 1000 of networks of given architectures) is trained  to find the configuration of the weight parameters  for which the error function is at the local minimum. The best model, favoured by the data, maximizes the  evidence. It is the  probability distribution, which only partially depends    on the  error function. It contains the Occam' contribution\footnote{It may happen that  for the model with the highest evidence the error function is not at global minimum.} (Eqs. \ref{evidence_Occam_volume} and \ref{evidence_Occam_symmetr}), which penalizes too complex models and allows one to choose the fit with the best predictive power.

The idea of the BNN formalism is to distinguish the statistical model, describing the data, which is characterized by good predictive power. To verify this property we make an estimate of $R_{+/-}$ for the new measurements done by VEPP-3 experiment \cite{Gramolin:2011tr}, which were not included  in the BNN analysis.  It can be noticed that the BNN and the HM predictions are  consistent with the new data (see  Fig. \ref{Fig_positron_data_vs_loops}).  For the future tests we provide  reader   with our predictions of $R_{+/-}$ for two other kinematics, which are going to be explored by the Novosibirsk experiment, see  Table \ref{Tab_vepp3}.
\begin{table}[h]
\begin{tabular}{c|c|c|c|c}
\hline
& & & &\\
$Q^2$(GeV$^2$) & $0.23$  &  $1.43$ & $0.82$ &   $0.96$           \cr
$\varepsilon$ & $0.95$ & $0.50$ & $0.42$       & $0.29$ \cr
\hline
            & $1.009$    & $1.017$ & $1.037$  & $1.043$\\
BNN          & $\pm0.007$ & $\pm 0.010$ & $\pm 0.007$ & $\pm 0.008$ \cr
\hline
            &  & & & \\
$\Box(N)$   & 0.999  & 1.026  & 1.020 &  1.028 \cr
$\Box(N+P_{33})$   & 1.001 & 1.020 & 1.019  & 1.027\cr
\end{tabular}
\caption{$R_{+/-}$ predictions for  VEPP-3 experiment \cite{Gramolin:2011tr} computed within the BNN and HM approaches. In the latter  the model contains either elastic $(N)$ or elastic and resonance $P_{33}$ TPE contributions (fit II). \label{Tab_vepp3}}
\end{table}\\

\section{Summary}

\label{Sec_Summary}

The TPE correction  was computed within the hadronic model. For the hadronic intermediate states  the proton and $P_{33}(1232)$ resonance were considered. The electromagnetic proton form factor parameters were obtained from the global fit to the cross section data only. Two FF parametrizations were discussed, the sum of monopoles and dipoles.   The TPE $\Box(P_{33})$ contribution was computed taking  different form of transition vertex and form factors than discussed  previously. In particular two parametrizations of the FFs for the $N\to P_{33}$  were considered.

The main goal of this paper was to confront the  predictions of the TPE effect coming from  the hadronic model and the Bayesian analysis of the $ep$ scattering data. The latter was preformed  by applying the neural network framework. The BNN response was constrained by the assumption that the PT data is not sensitive to TPE effect. Hence this comparison provides also a quantitative verification of this assumption.

It was demonstrated that the BNN and the hadronic model predictions agree on quantitative level in a wide $Q^2$ range. In particular for $Q^2$ between $1$ GeV$^2$ and $3$ GeV$^2$ the TPE corrections resulting form both approaches are very similar. In the intermediate $Q^2$ range ($Q^2 >3$ GeV$^2$) the agreement is  on  $2\sigma$ level. It is the kinematical domain where the data is limited. Obviously it  affects the BNN predictions. On the other hand in this kinematical limit the applied hadronic model  can be questionable.

For  $Q^2$ between $0.01$ GeV$^2$ and  $0.8$ GeV$^2$ the BNN and hadronic model predictions are inconsistent. In this $Q^2$ range the assumption (\ref{assumption_ABGG_1}) does not work effectively.  Similar inconsistency appears when one compares with the ABGG predictions of the TPE.   Indeed, in the ABGG approach the assumption  (\ref{assumption_ABGG_1}) also played a crucial role in the analysis.

A next step to improve the BNN approach would be to replace the assumption (\ref{assumption_ABGG_1}) by a weaker statement, and enlarge the number of independent TPE functions from one to six. The main problem is that the PT data seems to be not informative enough about $\varepsilon$ dependence of the TPE  effect.

\section*{Acknowledgements}

We thank J. Sobczyk, J. \. Zmuda and D. Prorok for reading the manuscript as well as C. Juszczak for useful hints about software development and his comments to the paper.

We thank also J. Arrington for his remarks  on the previous version of the paper.

Part of the calculations has been carried out in Wroclaw Centre for
Networking and Supercomputing (\url{http://www.wcss.wroc.pl}),
grant No. 268.\\

\appendix

\section{Evaluation of  TPE Term}

\label{Appendix_A}
We consider the proton electromagnetic FFs of the form,
\begin{equation}
\label{general_ff_form}
F_{i}(t) =\sum_{k=1}^{L_i}\sum_{n_k^i=1}^{N_k^i}
\frac{f^{n_k^i}_{i}}{(t-M_{i,n_k}^2)^k}, i=1, 2.
\end{equation}
Every $k-th$ pole function can be written as a derivative,
\begin{equation}
\frac{1}{(n-1)!}
\left.
\partial_{x}^{n-1}\frac{1}{t-x} \right|_{x=M^2}
=
\frac{1}{(t-M^2)^n}
\end{equation}
We introduce a notation,
\begin{equation}
\mathcal{D}^n_{M} f(M^2)\equiv \frac{1}{(n-1)!} \left. \frac{\partial^n f(x^2)}{\partial (x^2)^n} \right|_{x^2=M^2}, \;\; n=1,2,...,
\end{equation}
where
$
\mathcal{D}^0_{M} f(M^2) \equiv f(M^2).
$
Then  the form factor is written in the form,
\begin{equation}
F_{i}(t) =\sum_{k=1}^{L_i}\sum_{n_k^i=1}^{N_k^i}
\mathcal{D}_{M_{n_k^i}}^{k-1} \frac{f^{n_k^i}_{i}}{t-M_{n_k^i}^2}.
\end{equation}

We decompose both $w_{N,\parallel}$ and $w_{N,\times}$ functions into four components,
\begin{equation}
w_N = \sum_{i=1}^2\sum^2_{j=1} w^{ij}_N.
\end{equation}
The $ij$ component, due to its form factor, reads
\begin{widetext}
\begin{equation}
w^{ij}_N =
\sum_{k=1}^{L_i}\sum_{n_k^i=1}^{N_k^i}
\sum_{l=1}^{L_j}\sum_{n_l^j=1}^{N_l^j}
f^{n_k^i}_{i} f^{n_l^j}_{j}
\mathcal{D}_{M_{n_k^i}}^{k-1}
\mathcal{D}_{M_{n_l^j}}^{l-1}
\mathcal{E}( L^{\alpha\mu\nu} \mathcal{H}^{N,ij}_{\alpha\mu\nu}  ,M_{n_k^i},M_{n_l^j},M_p),
\end{equation}
where the leptonic tensor $L^{\alpha\mu\nu}$ is given by  either
\begin{equation}
\label{leptonic_tensor_box_direct}
L^{\alpha\mu\nu}_\parallel  =
\mathrm{Tr}\left((\hat{k}+m)\gamma^\alpha (\hat{k'}+m) \gamma^\mu  (\hat{k'} - \hat{l}+m) \gamma^\nu \right)
\end{equation}
or
\begin{equation}
L^{\alpha\mu\nu}_\times
=
\mathrm{Tr}\left((\hat{k}+m)\gamma^\alpha (\hat{k'}+m) \gamma^\mu  (\hat{k'} - \hat{l}+m) \gamma^\nu \right)
\label{leptonic_tensor_box_exchange}.
\end{equation}
The hadronic tensor reads,
\begin{equation}
\label{hadronic_ij_N}
\mathcal{H}^{N,ij}_{\alpha\mu\nu}
 =
\mathrm{Tr}\left((\hat{p} + M_p) \Gamma_\alpha(-q) (\hat{p'} + M_p )\Gamma_\mu^i(-l)  (\hat{p'} + \hat{l} + M_p)  \Gamma_\nu^j(q+l) \right),
\end{equation}
\begin{equation}
\Gamma^1_\mu(l) \equiv \gamma_\mu ,
\quad
\Gamma^2_\mu(l) = \frac{i\sigma^{\mu\nu} l_\nu}{2M_p}.
\end{equation}
$\mathcal{E}(\mathcal{N} ,m_a,m_b,m_h)$ is the one-loop integral defined as,
\begin{equation}
\label{I_parallel}
\mathcal{E}_\parallel(\mathcal{N} ,m_a,m_b,m_h)
=
\int \frac{d^4 l}{(2\pi)^4}
\frac{\mathcal{N}}{[ l^2 -m_a^2][(q+l)^2 -m_b^2][(q+l)^2 + i\epsilon][ l^2 + i\epsilon] [(k'-l)^2 - m^2 + i \epsilon][(p'+l)^2 - m^2_h+ i \epsilon]}
\end{equation}
for direct box diagram, and
\begin{equation}
\label{I_times}
\mathcal{E}_\times
(\mathcal{N} ,m_a,m_b,m_h)
=
\int \frac{d^4 l}{(2\pi)^4} \frac{  \mathcal{N}}{\left[ l^2 -m_a^2\right]\left[ (q+l)^2 -m_b^2 \right]\left[ (q+l)^2 + i \epsilon \right]\left[ l^2 + i\epsilon\right][(k+l)^2 - m^2 + i \epsilon][(p'+l)^2 - m^2_h+ i \epsilon]}
\end{equation}
for exchange box diagram.

In the case of the $P_{33}$ intermediate state we proceed in the similar manner. For instance for the  $P_{33}(Full)$ model we have
\begin{equation}
w_{\Delta} = \sum_{i=1}^3\sum^3_{j=1} w^{ij}_{\Delta}.
\end{equation}
The components of hadronic tensor  read,
\begin{equation}
\label{hadronic_ij_D}
\mathcal{H}^{\Delta,ij}_{\alpha\mu\nu}
=
\mathrm{Tr}\left((\hat{p} + M_p) \Gamma_\alpha(-q) (\hat{p'}+ M_p) \Gamma_{\mu\xi,i}^{\Delta,in}(-l,p'+l)  \left[\hat{p'} + \hat{l} + M_\Delta\right] \Lambda^{\xi\eta}(p'+l)  \Gamma_{\eta\nu,j}^{\Delta,out}(q+l,p'+l) \right),
\end{equation}
\end{widetext}
where
\begin{equation}
\Gamma_{\mu\nu,i}^{\Delta,in(out)} \equiv \Gamma_{\mu\nu}^{\Delta,in(out)}( C_k^V \to \delta_{ki}, k=3,4,5)
\end{equation}
In the case of $P_{33}(full)$ model for  the resonance form factors have a general form,
\begin{equation}
C_i^V(t) =  - c_i^V  \frac{a_i M_V^2 }{ t - a_i M_V^2}\frac{M_V^4}{(t-M_V^2)^2}.
\end{equation}
They can be written in the form,
\begin{equation}
C_i^V(t) = \frac{c_{i}^{1,1}}{\left[t-(M_{i}^{1,1})^2\right]^2}+ \frac{c_{i}^{1,2}}{\left[t-(M_i^{1,2})^2\right]}  + \frac{c_{i}^{2,1}}{\left[t-(M_{i}^{2,1})^2\right]^2},
\end{equation}
where
\begin{eqnarray}
C_i^{1,1} &=&  - c_i^V    \frac{a_i M_V^2}{(a_i-1)^2}, \quad M_i^{1,1}=\sqrt{a_i} M_V,\\
C_i^{1,2} &=&   c_i^V    \frac{a_i M_V^2}{(a_i-1)^2}, \quad M_i^{1,1}= M_V \\
C_i^{2,1} &=&   c_i^V    \frac{a_i M_V^4}{(a_i-1)}, \quad M_i^{1,1}= M_V.
\end{eqnarray}

\begin{widetext}

\begin{equation}
w^{ij}_\Delta =
\sum_{k=1}^{2}
\sum_{n_k=1}^{N_k}
\sum_{l=1}^{2}
\sum_{n_l=1}^{N_l}
c^{k, n_k}_{i} c^{l, n_l}_{j}
\mathcal{D}_{M_i^{k,n_k}}^{k-1}
\mathcal{D}_{M_j^{l,n_l}}^{l-1}
\mathcal{I}( L^{\alpha\mu\nu} \mathcal{H}^{\Delta,ij}_{\alpha\mu\nu}  ,M_i^{k,n_k},M_j^{l,n_l},M_\Delta),
\end{equation}
where
$N_1=2$, $N_2=1$.
\end{widetext}

The algebraical calculations, like  computing the leptonic (\ref{leptonic_tensor_box_direct}), (\ref{leptonic_tensor_box_exchange}) and hadronic  (\ref{hadronic_ij_N}), (\ref{hadronic_ij_D}) tensors and their contractions, are done with help of  FeynCalc package \cite{Mertig:1990an,Mertig:1998vk}.

The integrals (\ref{I_parallel}), (\ref{I_times}) are expressed (also with the help of routines in FeynCalc) in terms of Veltman-Passariono scalar loop integrals \cite{tHooft_1979}\cite{Passarino79}. Because of the complex structure of the the numerators of the integrals (\ref{I_parallel}), (\ref{I_times})   some  pre-reduction  the numerator with  the denominator is necessary.

Having the analytic expressions for all integral components, their numerical values are computed with  LoopTool library \cite{vanOldenborgh:1989wn,Hahn:1998yk}.

\section{$\chi^2$}

\label{Appendix_B}

In order to get the FFs parameters of the parametrization I (\ref{FF_polesum}) and II (\ref{FF_dipolesum}) we analyse the same   unpolarized cross section data as  in Ref. \cite{Graczyk:2011kh}.

We consider the following $\chi^2$ function,
\begin{equation}
\label{chi2}
\chi^2 = \sum_{k=1}^{N} \left[ \sum_{i=1}^{n_k} \left(\frac{\lambda_k \sigma^{th}_{ki} - \sigma^{ex}_{ki}}{\Delta\sigma_{ki}}\right)^2 +
\left(\frac{\lambda_{k}-1}{\Delta \lambda_k } \right)^2 \right].
\end{equation}
$N = 28$ is a number of independent data sets in the fit; $n_k$ is a number of points in the $k$-th data set; $\sigma_{ki}^{th}$ is the reduced cross section given by Eq. (\ref{sigma_reduced_2photon}), while $\sigma^{ex}_{ki}$ and $\Delta\sigma^{ex}_{ki}$ denote the experimental measurement and its error. By $\Delta \lambda_k$'s the systematic normalization errors  are introduced. For every data set the normalization parameter $\lambda_k$ is established from the fit. A treatment of the systematic normalization errors  is the same as in Refs. \cite{Arrington:2003df,Alberico:2008sz,Graczyk:2011kh,Arrington:2007ux}. For the statistical  explanation of this procedure see Ref. \cite{D'Agostini:1995fv}.

\end{document}